\documentclass[fleqn,usenatbib]{mnras}

\usepackage{newtxtext,newtxmath}

\usepackage[T1]{fontenc}
\usepackage{ae,aecompl}

\usepackage{graphicx}	
\usepackage{amsmath}	
\usepackage{amssymb}	
\usepackage{threeparttable}
\newcommand\ktwo{{\it K2}}
\newcommand\kepler{{\it Kepler}}
\newcommand\corot{{\it CoRoT}}
\newcommand\gaia{{\it Gaia}}
\newcommand\Mjup{M$_\mathrm{Jup}$} 
\newcommand\Rjup{R$_\mathrm{Jup}$}
\newcommand\kms{km\,s$^{-1}$} 
\newcommand\gcm{g\,cm$^{-3}$} 
\newcommand\teff{$T_{\rm eff}$}
\newcommand\logg{log\,{\it g$_\star$}}
\newcommand\vmic{$v_{\rm mic}$}
\newcommand\vsini{$v$\,sin\,$i_\star$}
\newcommand\vmac{$v_{\rm mac}$}

\newcommand\Prot{$P_{\mathrm{rot}}$}
\newcommand\Msun{$\mathrm{M_{\sun}}$}  
\newcommand\Rsun{$\mathrm{R_{\sun}}$} 
\newcommand\Mearth{$\mathrm{M_{\earth}}$}  
\newcommand\Rearth{$\mathrm{R_{\earth}}$}
\newcommand\ms{m\,s$^{-1}$}

\title[K2-140b and K2-180b]{K2-140b and K2-180b --  Characterization of a hot Jupiter and a mini-Neptune from the \ktwo\ mission }

\author[J. Korth et al.]{
J.~Korth$^{1}$\thanks{E-mail: Korthj@uni-koeln.de},
Sz.~Csizmadia$^{2}$,
D.~Gandolfi$^{3}$,
M.~Fridlund$^{4,5}$,
M.~P\"atzold$^{1}$,
T.~Hirano$^{6}$,
\newauthor
J.~Livingston$^{7}$,
C.~M.~Persson$^{4}$,
H.~J.~Deeg$^{8,9}$,
A.~B.~Justesen$^{10}$,
O.~Barrag\'an$^{3}$,
S.~Grziwa$^{1}$,
\newauthor
M.~Endl$^{11}$,
R.~Tronsgaard$^{12,13}$,
F.~Dai$^{14,15}$,
W.~D.~Cochran$^{11}$,
S.~Albrecht$^{10}$,
R.~Alonso$^{8,9}$,
\newauthor
J.~Cabrera$^{2}$,
P.~W.~Cauley$^{16}$,
F.~Cusano$^{17}$,
Ph.~Eigm\"uller$^{2,18}$,
A.~Erikson$^{2}$,
M.~Esposito$^{19}$,
\newauthor
E.~W.~Guenther$^{19}$,
A.~P.~Hatzes$^{19}$,
D.~Hidalgo$^{8,9}$,
M.~Kuzuhara$^{20,21}$,
P.~Monta\~{n}es$^{8,9}$,
\newauthor
N.~R.~Napolitano$^{22}$,
N.~Narita$^{7,9,20,21}$,
P.~Niraula$^{16}$,
D.~Nespral$^{8,9}$,
G.~Nowak$^{8,9}$,
E.~Palle$^{8,9}$,
\newauthor
C.~E.~Petrillo$^{23}$,
S.~Redfield$^{16}$,
J.~Prieto-Arranz$^{8,9}$,
H.~Rauer$^{2,18,24}$,
A.~M.~S.~Smith$^{2}$,
\newauthor
C.~Tortora$^{23}$,
V.~Van~Eylen$^{5}$,
J.~N.~Winn$^{15}$
\\
$^{1}$Rheinisches Institut f\"ur Umweltforschung an der Universit\"at zu K\"oln, Abteilung Planetenforschung, Aachener Str. 209, 
50931 K\"oln, Germany\\
$^{2}$Institute of Planetary Research, German Aerospace Center, Rutherfordstrasse 2, 12489 Berlin, Germany\\
$^{3}$Dipartimento di Fisica, Universit\`a di Torino, Via P. Giuria 1, I-10125, Torino, Italy\\
$^{4}$Department of Space, Earth and Environment, Chalmers University of Technology, Onsala Space Observatory, 439 92 Onsala, Sweden\\
$^{5}$Leiden Observatory, University of Leiden, PO Box 9513, 2300 RA, Leiden, The Netherlands\\
$^{6}$Department of Earth and Planetary Sciences, Tokyo Institute of Technology, 2-12-1 Ookayama, Meguro-ku, Tokyo 152-8551, Japan\\
$^{7}$Department of Astronomy, Graduate School of Science, The University of Tokyo, Hongo 7-3-1, Bunkyo-ku, Tokyo, 113-0033, Japan\\
$^{8}$Departamento de Astrof\'isica, Universidad de La Laguna, E-38206, Tenerife, Spain\\
$^{9}$Instituto de Astrof\'isica de Canarias, E-38205, La Laguna, Tenerife, Spain\\
$^{10}$Stellar Astrophysics Centre, Deparment of Physics and Astronomy, Aarhus University, Ny Munkegrade 120, DK-8000 Aarhus C, Denmark\\
$^{11}$Department of Astronomy and McDonald Observatory, University of Texas at Austin, 2515 Speedway, Stop C1400, Austin, TX 78712, USA\\
$^{12}$Nordic Optical Telescope, Rambla Jos\'e Ana Fern\'andez P\'erez 7, 38711 Bre\~na Baja, Spain\\
$^{13}$DTU Space, National Space Institute, Technical University of Denmark, Elektrovej 328, DK-2800 Kgs. Lyngby, Denmark\\
$^{14}$Department of Physics and Kavli Institute for Astrophysics and Space Research, Massachusetts Institute of Technology,\\
Cambridge, MA, 02139, USA\\
$^{15}$Department of Astrophysical Sciences, Princeton University, 4 Ivy Lane, Princeton, NJ, 08544, USA\\
$^{16}$Astronomy Department and Van Vleck Observatory, Wesleyan University, Middletown, CT 06459, USA\\
$^{17}$INAF - Osservatorio di Astrofisica e Scienza dello Spazio di Bologna, Via Gobetti 93/3, I-40129 Bologna, Italy\\
$^{18}$Center for Astronomy and Astrophysics, TU Berlin, Hardenbergstr. 36, 10623 Berlin, Germany\\
$^{19}$Th\"uringer Landessternwarte Tautenburg, Sternwarte 5, D-07778 Tautenberg, Germany\\
$^{20}$Astrobiology Center, NINS, 2-21-1 Osawa, Mitaka, Tokyo 181-8588, Japan\\
$^{21}$National Astronomical Observatory of Japan, NINS, 2-21-1 Osawa, Mitaka, Tokyo 181-8588, Japan\\
$^{22}$INAF - Osservatorio Astronomico di Capodimonte, Salita Moiariello, 16, I-80131 Napoli, Italy\\
$^{23}$Kapteyn Astronomical Institute, University of Groningen, Postbus 800, 9700 AV, Groningen, The Netherlands\\
$^{24}$Institut f\"ur Geologische Wissenschaften, Freie Universit\"at Berlin, Malteserstr. 74-100, 12249 Berlin, Germany
}
\date{Accepted XXX. Received YYY; in original form ZZZ}

\pubyear{2018}

\begin{document}
\label{firstpage}
\pagerange{\pageref{firstpage}--\pageref{lastpage}}
\maketitle

\begin{abstract}
We report the independent discovery and characterization of two \ktwo\ planets: K2-180b, a mini-Neptune-size planet in an 8.9-day orbit transiting a V\,=\,12.6\,mag, metal-poor ([Fe/H]\,=\,$-0.65\pm0.10$) K2V star in \ktwo\ campaign 5; K2-140b, a transiting hot Jupiter in a 6.6-day orbit around a V\,=\,12.6\,mag G6V ([Fe/H]\,=\,$+0.10\pm0.10$) star in \ktwo\ campaign 10. Our results are based on \ktwo\ time-series photometry combined with high-spatial resolution imaging and high-precision radial velocity measurements. We present the first mass measurement of K2-180b. K2-180b has a mass of $M_\mathrm{p}=11.3\pm1.9$\,\Mearth\ and a radius of $R_\mathrm{p}=2.2\pm0.1$\,\Rearth, yielding a mean density of $\rho_\mathrm{p}=5.6\pm1.9$\,\gcm, suggesting a rock composition. Given its radius, K2-180b is above the region of the so-called ``planetary radius gap''. K2-180b is in addition not only one of the densest mini-Neptune-size planets, but also one of the few mini-Neptune-size planets known to transit a metal-poor star. We also constrain the planetary and orbital parameters of K2-140b and show that, given the currently available Doppler measurements, the eccentricity is consistent with zero, contrary to the results of a previous study.
\end{abstract}



\section{Introduction}

One of the most astonishing results from the study of planets orbiting stars other than the Sun is the variety of exoplanetary systems \citep{Hatzes2016}. Gas-giant planets with orbital periods shorter than $\sim$10\,days (the so-called hot Jupiters), as well as small planets with radii between $\sim$1.5 and 4\,\Rearth\ (super-Earths and mini-Neptunes) established new groups of planets that are not present in our Solar System \citep[see, e.g.,][]{Mayor1995,Leger2009}. 

Those small exoplanets, mostly detected by the \kepler\ mission\footnote{The project HARPS and ETAEARTH are also focusing on small exoplanets \citep[e.g.][]{Dumusque_et_al_2012,pepe_et_al_2013}.}, permit the study of the occurrence rate of small planets for the first time \citep[e.g.][]{Burke_et_al_2015}. By studying the planetary distributions the so-called "planetary radius gap" was discovered. The planetary radius distribution for short-period planets seems to be bimodal with a lack of planets between 1.5 and 2\,\Rearth\,\citep{fulton_et_al_2017,Van_Eylen_et_al_2018}. The gap had been predicted by photoevaporation models \citep[e.g.][]{lopez_fortney_2013,Owen_wu_2013,Jin_et_al_2014,Owen_wu_2017,Jin_Mordasini_2018} wherein the planet may lose its atmosphere due to stellar radiation. Therefore the gap separates planets with (>\,2\,\Rearth) and without gaseous envelopes (<\,1.5\,\Rearth). \citet{Ginzburg_et_al_2018} suggested another mechanism in which the luminosity of a cooling core activates the mass loss. In a recently published study, \citet{Fulton_petigura_2018} found evidence for photoevaporation, but could not exclude the possibility that both mechanisms are operative. \citet{Fulton_petigura_2018} also figured out that the location of the radius gap is dependent on the stellar mass.

Another relevant dependence of the planetary distribution is the stellar metallicity which was studied by e.g. \citet{Mortier_et_al_2012,Wang_fischer_2015,Mortier_et_al_2016,Buchhave_et_al_2018,Petigura_et_al_2018}. Stellar metallicity is a key parameter for understanding the evolution and formation of planetary systems \citep[e.g][]{Buchhave_et_al_2014}. While \citet{Mortier_et_al_2012} found a correlation between planetary mass and host star's metallicity for gas giants, the correlation for smaller planets is still investigated \citep[e.g][]{Wang_fischer_2015,Mortier_et_al_2016}. The correlation between the occurrence rate and the metallicity of the host star for Neptune-like planets seems to be weakest \citep{Courcol_et_al_2016}. However, close-in exoplanets (P\,<\,10\,days) are found to be more common around metal-rich stars with an excess of hot rocky planets \citep{Mulders_et_al_2016} and of hot Neptunes \citep{Dong_et_al_2018}. \citet{petigura_et_al_2018b} also pointed out that planetary occurrence and stellar metallicity are not correlated for every planetary size and orbital period. The overall finding of their study, that there exists a great diversity around metal-rich stars, corroborated that planets larger than Neptune are more common around metal-rich stars, while planets smaller than Neptune exist around stars with different metallicities. In a recently published paper, \citet{owen_murray-clay_2018} studied the connection between stellar metallicity dependency of planetary properties, like the orbital period and planetary size. They investigated also how the location of the planetary radius gap and its possible source, photoevaporation, for close-in, low-mass planets dependent on the stellar metallicity. One of their main outcomes was that solid core masses of planets are larger around metal-rich stars and that these cores are able to accrete larger gaseous envelopes \citep{owen_murray-clay_2018}.  

An extraordinary diversity exists not only in the mass-radius parameter space, but also in the architecture of exoplanetary systems \citep{Winn2015}. This diversity still lacks a complete theoretical understanding. It is therefore important to continue to increase the exoplanet database using data of improved accuracy to provide input to modeling efforts. Although many exoplanets have been discovered so far ($\sim$3800, as of September 2018\footnote{\url{http://exoplanet.eu/catalog/}.}), only a small fraction of objects have a precise radius and mass measurements that allow the deviation of their internal compositions \citep{valencia_et_al_2007,Wagner_et_al_2011}. In particular, precise mass and radius measurements (better than 20\,\% in mass and radius) are needed to distinguish between various possible planetary compositions\footnote{ In the future, accuracies up to 2\,\%, 4-10\,\% and 10\,\% in stellar radii, masses and ages are achievable with \textit{PLATO}, respectively \citep{Rauer_et_al_2014}.}. High signal-to-noise ratio data can only be collected by observing bright host stars (V\,<\,13\,mag) from ground, as well as from space. 

The \ktwo\ mission \citep{howell_et_al_2014} and the \textit{TESS} mission \citep{Ricker_et_al_2014} are currently the only surveys that search for transiting exoplanets from space. \ktwo\ is discovering planets orbiting stars that are on average 2--3 magnitudes brighter than those targeted by the original \kepler\ mission \citep[e.g.][]{crossfield_et_al_2016}. These bright stars are located in different fields (designated ``campaigns'') along the ecliptic. The space telescope is re-targeted every $\sim$80\,days. While \ktwo\ transit light curves (LC) provide the relative planetary radii $R_\mathrm{p}/R_\star$, planetary masses can be determined through ground-based radial velocity (RV) follow-up observations. The quality of the ground-based high-resolution spectroscopy and RV measurements are significantly improved since the stars are almost exclusively brighter than those hitherto observed by the \kepler\ mission. \citet{vanderburg_johnson_2014} give for a V\,=\,12\,mag star a photometric precision of $\sim$30\,ppm. 

K2-180 and K2-140 are two stars that were observed by \ktwo\ during campaign 5 and 10 (C5 and C10), respectively. Each star was found to host a transiting planet: K2-180b, a mini-Neptune-size planet candidate which was first reported by \citet{pope_et_al_2016} and recently validated as a planet by \citet{mayo_et_al_2018} without any mass determination; K2-140b, a hot Jupiter in a 6.57-day orbit, which was recently discovered and confirmed by \citet{Giles_et_al_2018} (hereafter G18) and also statistically validated by \citet{Livingston_et_al_2018} as well as in \citet{mayo_et_al_2018}.
 
In this paper, the KESPRINT team\footnote{\url{http://www.iac.es/proyecto/kesprint/}\\The KESPRINT team merged from two teams: the "K2 Exoplanet Science Team" (KEST) \citep{Johnson2016} and the "Equipo de Seguimiento de Planetas Rocosos Intepretando sus Transitos'' (ESPRINT) \citep{Sanchis-Ojeda2015} team.} combines the \ktwo\ photometry with ground-based high-resolution imaging and high-precision RV measurements in order to confirm the planetary nature of K2-180b, as well as to characterize independently K2-140b. Both planetary systems are found here to be well characterized including the planetary masses, sizes, and bulk densities. For K2-180b, the first mass measurement is reported. K2-180b is of particular interest not only because of its radius ($R_\mathrm{p}=2.2 \pm 0.1$\,\Rearth) which is slightly above the planetary radius gap, but also because of its host star's metallicity. K2-180b is one of a few mini-Neptunes orbiting a metal-poor star. The K2-140b's RV measurements presented in this paper doubled the number of existing Doppler measurements for this star, allowing studies of the non-zero eccentricity claimed by G18.

\section{Observations}
\label{sec:observation}
\subsection{\ktwo\ photometry and transit detection}
\label{sec:photometry}

\begin{table}
\centering
 \caption{Main identifiers, equatorial coordinates, selected magnitudes and proper motion, and parallax of K2-180 and K2-140.} 
 \label{tab:stellartab}
 \begin{tabular}{lcc}
 \hline
 \hline
 \noalign{\smallskip}
  \, & K2-180 & K2-140 \\
  \noalign{\smallskip}
  \hline
  \noalign{\smallskip}
  \multicolumn{3}{l}{Main identifiers} \\
  \noalign{\smallskip}
  EPIC ID$^{(\mathrm{a})}$  & 211319617 & 228735255 \\
  \gaia\ ID$^{(\mathrm{b})}$ & 600750922666388992 & 3579426053724051584\\
  2MASS ID$^{(\mathrm{a})}$ & 08255135+1014491 & 12323296-0936274\\
  UCAC2 ID$^{(\mathrm{c})}$ & 201-069327 & 161-076473\\
  UCAC4 ID$^{(\mathrm{a})}$ & 502-048219 & 402-053388\\
  \noalign{\smallskip}
  \hline
  \noalign{\smallskip}
   \multicolumn{3}{l}{Equatorial coordinates [J2000.0]$^{(\mathrm{d})}$ }\\
   \noalign{\smallskip}
  $\alpha$   & $08^{h}25^{m}51^{s}.35$& $12^{h}32^{m}32^{s}.96$\\
  $\delta$  & 10\degr14\arcmin49\arcsec.13 & -09\degr36\arcmin27\arcsec44\\
  \noalign{\smallskip}
  \hline
  \noalign{\smallskip}
  \multicolumn{3}{l}{Apparent magnitudes [mag]}\\
  \noalign{\smallskip}
  B$^{(\mathrm{a})}$ & 13.334 $\pm$ 0.010 & 13.349 $\pm$ 0.030\\
  V$^{(\mathrm{a})}$ & 12.601 $\pm$ 0.020 & 12.624 $\pm$ 0.030\\
  J$^{(\mathrm{d})}$ & 11.146 $\pm$ 0.023 & 11.421 $\pm$ 0.026\\
  H$^{(\mathrm{d})}$ & 10.747 $\pm$ 0.026 & 11.068 $\pm$ 0.021\\
  Ks$^{(\mathrm{d})}$ & 10.677 $\pm$ 0.026 & 10.995 $\pm$ 0.021\\
  g$^{(\mathrm{a})}$ & 12.900 $\pm$ 0.020 & 12.930 $\pm$ 0.060\\
  r$^{(\mathrm{a})}$ & 12.376 $\pm$ 0.020 & 12.426 $\pm$ 0.020\\
  i$^{(\mathrm{a})}$ & 12.176 $\pm$ 0.020 & 12.292 $\pm$ 0.050\\
  \noalign{\smallskip}
  \hline
  \noalign{\smallskip}
  \multicolumn{3}{l}{Proper motion [mas yr$^{-1}$]$^{(\mathrm{c})}$ and parallax [mas]$^{(\mathrm{b})}$}\\ 
  \noalign{\smallskip}
  $\mu_{\alpha}$ cos$\delta$  & 97.8 $\pm$ 1.9 & -0.4 $\pm$ 2.4\\
  $\mu_{\delta}$ & -84.8 $\pm$ 1.3 & -2.1 $\pm$ 2.5\\
  parallax $p$				 & 4.88 $\pm$ 0.11 &  2.85 $\pm$ 0.12 \\
  \noalign{\smallskip}
  \hline
  \hline
  \end{tabular}
  \begin{tablenotes}\footnotesize
  \item Taken from $^{(\mathrm{a})}$ Ecliptic Planet Input Catalog (\url{http://archive.stsci.edu/k2/epic/search.php}),$^{(\mathrm{b})}$ \gaia\ archive \citep{gaia_mission_2016,brown_et_al_gaia_2018}, $^{(\mathrm{c})}$ UCAC4 \citep{Zacharias_et_al_2012} and $^{(\mathrm{d})}$ 2MASS \citep{Cutri_et_al_2003,Skrutskie_et_al_2006}. 
  \end{tablenotes}
\end{table}

K2-180, EPIC\,211319617 (Table~\ref{tab:stellartab}), was observed by the \ktwo\ mission during C5, between 2015 April 15 and 2015 July 10. It was proposed by programs GO5007, G05029, G05060, and G05106\footnote{\url{https://keplerscience.arc.nasa.gov/k2fields.html} The proposers of the individual programs are J.~N.~Winn (G05007), D.~Carbonneau (G05029), J.~Coughlin (G05060, and B.~Jackson (G05106).}. The telescope's field-of-view (FoV) was centered at coordinates $\alpha=08^{h}25^{m}51^{s}.35$, $\delta=10\degr14\arcmin49\arcsec.13$. 

K2-140, EPIC\,228735255 (Table~\ref{tab:stellartab}), was observed during \ktwo\,'s C10 between 2016 July 06 and 2016 September 20, and was proposed by programs GO10068 and GO10077\footnote{\url{https://keplerscience.arc.nasa.gov/k2-fields.html}. The proposers of the individual programs are D.~Charbonneau (G010068) and A.~Howard (G010077).}. The telescope FoV was pointed towards coordinates $\alpha=12^{h}32^{m}32^{s}.96$, $\delta=-09\degr36\arcmin27\arcsec.44$. A 3.5-pixel initial pointing error which occurred at the beginning of C10, was corrected after six days. The data were separated into two segments. The loss of module 4 on 2017 July 20 resulted in a data gap of 14 days.

Different algorithms are used by KESPRINT for the detection of transit-like signals in time-series photometric data. The detection algorithms D\'etection Sp\'ecialis\'ee de Transits (DST) from DLR Berlin \citep{cabrera_et_al_2012} and EXOTRANS from RIU-PF Cologne \citep{grziwa_et_al_2012} were applied to the data of C5 and C10 that were pre-processed by \citet{vanderburg_johnson_2014}. Light curves were also extracted from the calibrated data following the procedures described by \citet{Dai_et_al_2017} at MIT/Princeton. Briefly, the target pixel files were downloaded from the Mikulski Archive for Space Telescopes and were converted to light curves by a similar approach described by \citet{vanderburg_johnson_2014}. Circular apertures are placed around the brightest pixel within the postage stamp and its radius is varied according to the Kepler magnitude of the target so that brighter target stars have larger apertures. The intensity fluctuations due to the rolling motion of the spacecraft are identified by fitting a 2-D Gaussian function to the aperture-summed flux distribution. A piecewise linear function is fitted between the flux variation and the centroid motion of the target which is afterward detrended from the observed flux variation to produce a light curve. 

\begin{figure}
\includegraphics[width=\columnwidth]{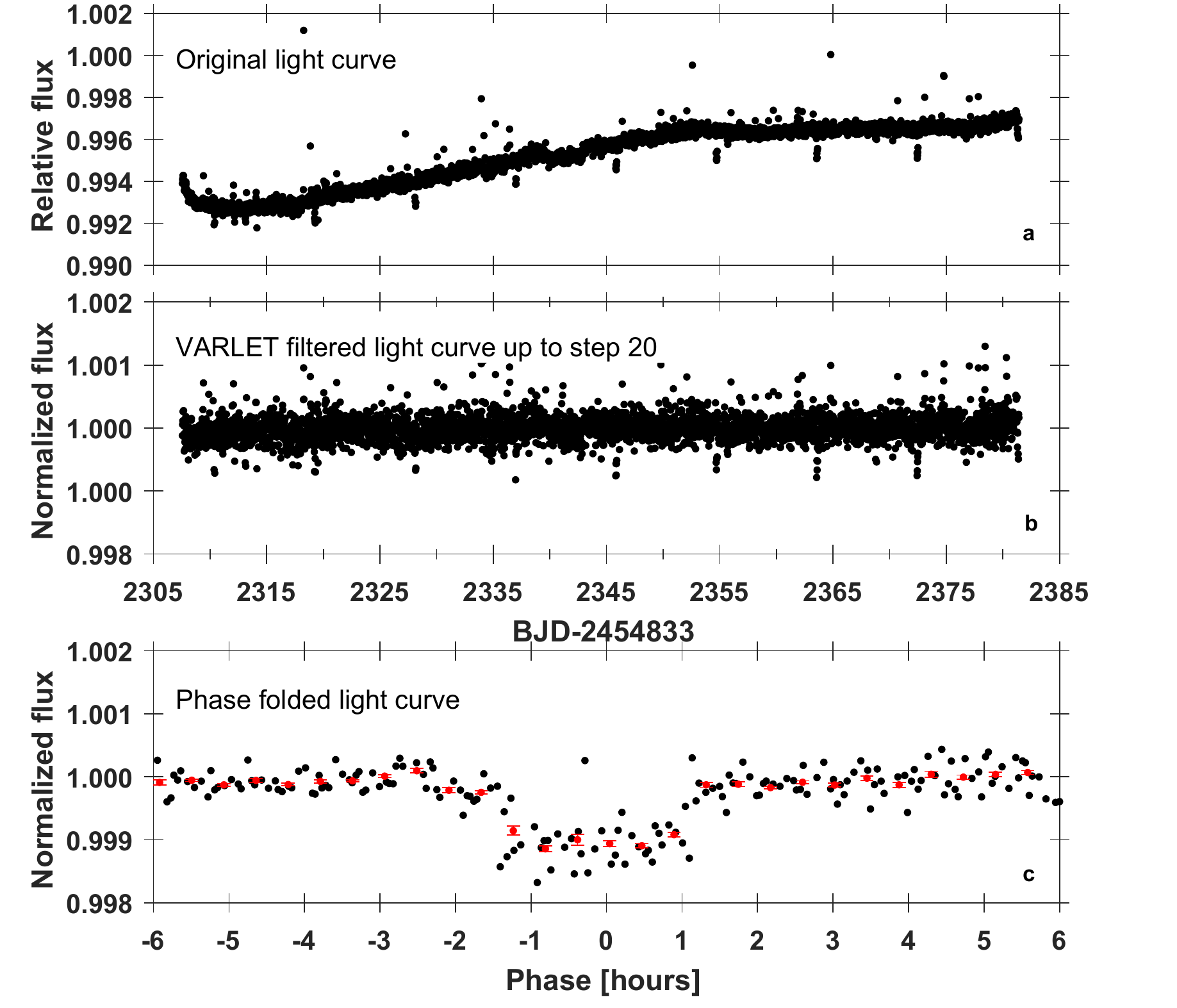}
\caption{(a) The original light curve of K2-180 from \citet{vanderburg_johnson_2014}; (b) VARLET filtered up to step 20 containing a periodic signal with a period of 8.87~days; (c) phase folded (black) and binned (red) light curves with a binning of 0.002. The changed transit depth in the VARLET filtered light curve (b) is clearly visible.}
\label{fig:C5_9617_detection}
\end{figure}

RIU-PF filters the light curves using the wavelet-based filter VARLET \citep{grziwa_paetzold_2016} prior to the transit search in order to reduce stellar variability and instrument systematics. VARLET allows a different strength of filtering. An example of a low level of filtering is shown in Fig.~\ref{fig:C5_9617_detection} (panel~b). This reduces substantially the stellar variability and instrumental systematics. The selected filtering level leads to a shallower transit depth which has, however, no influence on the detection efficiency of EXOTRANS. 

This code, as well as the code developed by \citet{Dai_et_al_2017}, uses a modification of the Box-Least-Squared (BLS) algorithm \citep{kovacs_et_al_2002,Ofir_2014} to search for periodic signals. DST uses an optimized transit shape, with the same number of free parameters as BLS, and an optimized statistic for signal detection. The algorithm in EXOTRANS changes the box size (transit duration) as a function of the searched orbital period by also taking, if available, the stellar radius into account. 

If a periodic transit signal is detected by EXOTRANS, a second filter, PHALET \citep{grziwa_paetzold_2016} that combines wavelets with phase-folding of well-known periods, removes this transit at the detected period and the light curve is searched again by EXOTRANS. This procedure is repeated until a certain signal detection efficiency (SDE) value is achieved. For every detected period a SDE value is calculated. This SDE value is compared to a SDE threshold. This SDE threshold was empirically estimated and is 6 for the \ktwo\ mission. If this threshold is not achieved the search stops after 15 iterations to save computer time. This automation allows one to search for additional transit-like signals in the stellar light curve. An additional check of the detected periodic signals is implemented by comparing all detected periods and phases. Most of the background binaries are also removed this way. After this procedure, an overall SDE threshold is calculated using a Generalized Extreme Value (GEV) distribution. If a LC has a SDE value above this threshold the LC is sorted out for further inspections and investigations.

The use of independent detection algorithms and different filter techniques maximizes the confidence in transit detections as well as decreases the number of false positive detections \citep{Moutou_et_al_2005}. This approach was successfully used for the search in \corot\ and \kepler\ light curves and is also used within the KESPRINT team for the detection and confirmation of planetary candidates from the \ktwo\ mission \citep[e.g.][]{grziwa_et_al_2016,Niraula_et_al_2017,Hirano_et_al_2018} and \textit{TESS} mission.

\begin{figure}
\includegraphics[width=\columnwidth]{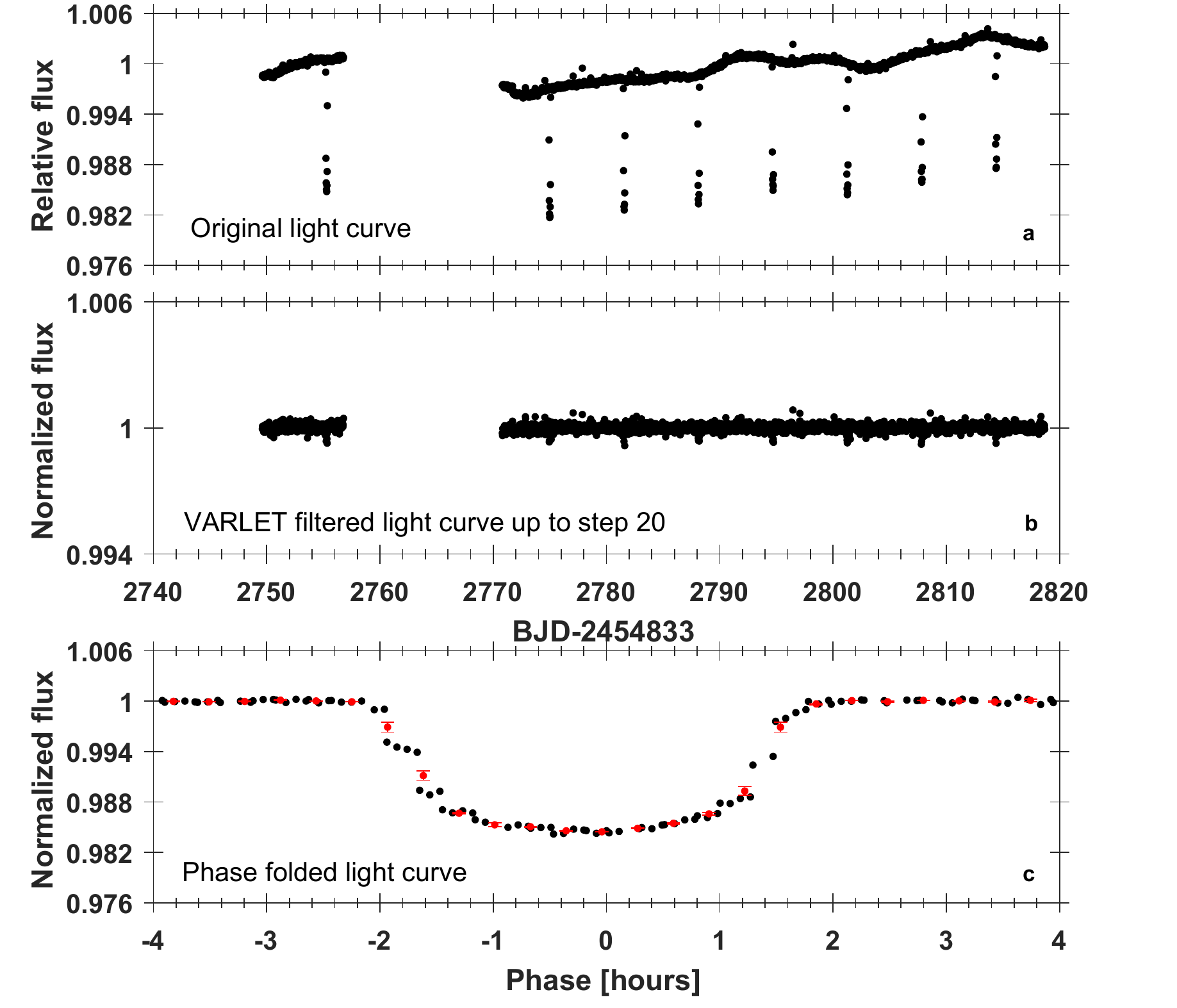}
\caption{(a) The original light curve of K2-140 from \citet{vanderburg_johnson_2014}; (b) VARLET filtered up to step 20 containing a periodic signal with a period of 6.57~days; (c) phase folded (black) and the binned (red) light curve with a binning of 0.002. Note that the phase folding with only a first guess on the orbital period leads to the arrangement of the individual observation points.The changed transit depth in the VARLET filtered light curve (b) is clearly visible.}
\label{fig:C10_5255_detection}
\end{figure}

All three methods detected transit-like signals in the light curves of K2-180 and K2-140 at a period of 8.87~days with a depth of 0.12~\% and 6.57~days with a depth of 1.6~\%, respectively (panel~c in Fig.~\ref{fig:C5_9617_detection} and Fig.~\ref{fig:C10_5255_detection}). 

To further exclude a contaminating scenario by a background binary, the even/odd differences were computed, which show no depth difference within 1$\sigma$. Also, no secondary eclipses were found at phases 0.5.

\subsection{Ground-based follow-up observations} 
\label{sec:Ground_Based_FU}
Ground-based, high-spatial resolution imaging of K2-180 and K2-140 was performed with the NASA Exoplanet Star and Speckle Imager (NESSI) and with the Infrared Camera and Spectrograph (IRCS) with adaptive optics (AO) to exclude the presence of potentially unresolved binaries and rule out false positive scenarios. Additionally, seeing-limited observations with the Andalucia Faint Object Spectrograph and Camera (ALFOSC) observations of K2-180 were carried out to measure the light contamination factor arising from nearby sources whose light leaks into the photometric mask of the target. In order to confirm the planetary nature of the transit signals, derive the fundamental stellar parameters, and measure the masses of the two planets, high-precision RV follow-up observations of both stars were secured with the FIbre-fed \'Echelle Spectrograph (FIES). K2-180 was also observed with the HARPS-N spectrograph. A description of the ground-based follow-up observations is provided in the following subsections. 

\subsubsection{NESSI speckle imaging}
\label{sec:speckle}
\begin{figure*}
\includegraphics[width=\columnwidth]{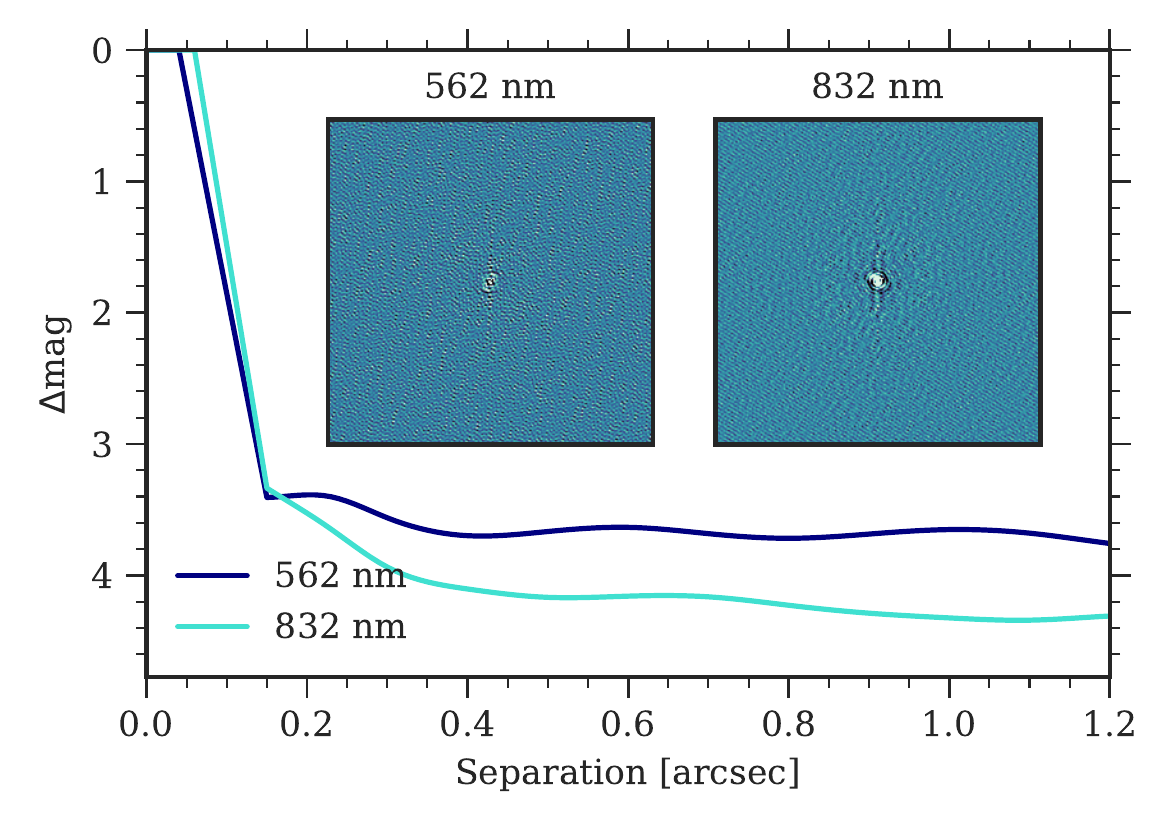}
\includegraphics[width=\columnwidth]{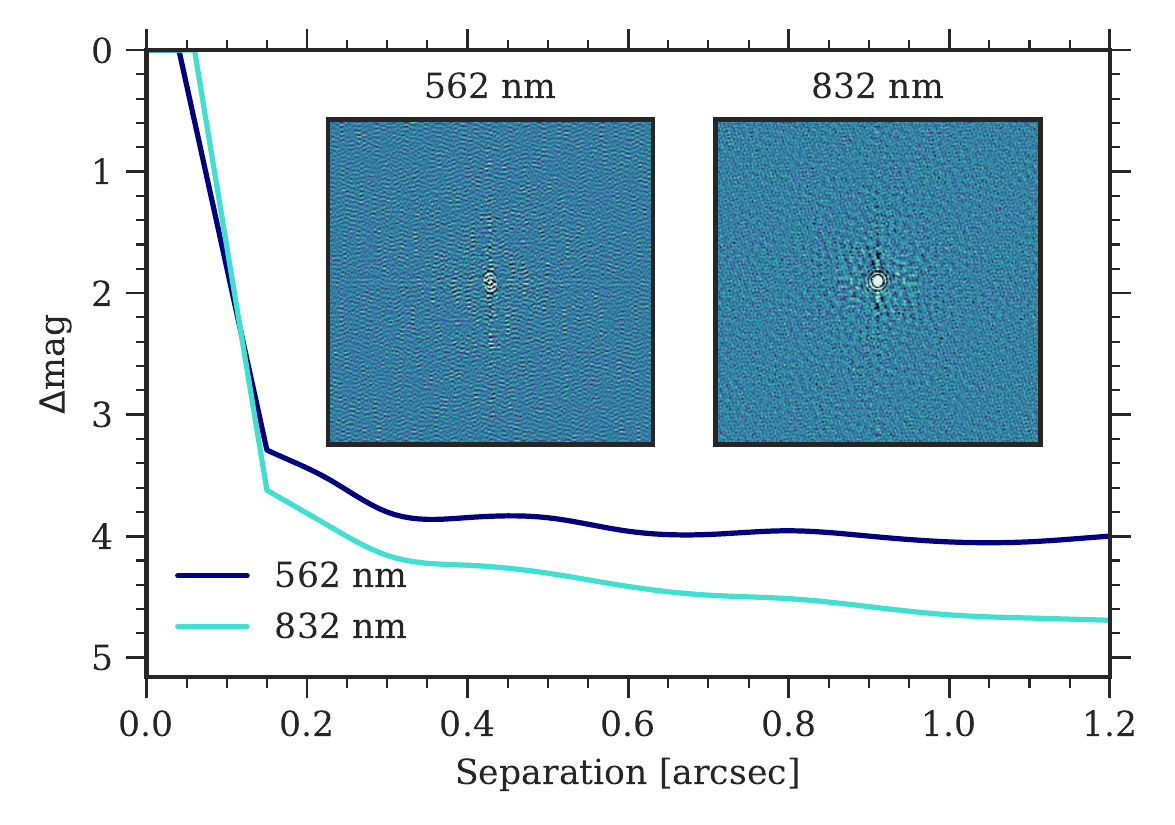}
\caption{5$\sigma$ contrast curves based on the NESSI speckle imaging for K2-180 (left) and K2-140 (right). The blue and light blue curves are the blue band centered at 562\,nm with a width of 44\,nm and the red band centered at 832\,nm with a width of 40\,nm, respectively. The insets display $4.6\arcsec\times4.6\arcsec$ images of each star.}
\label{fig:speckle_fig}
\end{figure*}

Both K2-140 and K2-180 were observed with NESSI on the 3.5\,m WIYN telescope at the Kitt Peak National Observatory, Arizona, USA on the nights of 2017 March 10 and 2017 May 11, respectively. NESSI is a new instrument that uses high-speed electron-multiplying CCDs (EMCCDs) to capture sequences of 40 ms exposures simultaneously in two bands \citep{2016SPIE.9907E..2RS}. In addition to the target, nearby point source calibrator stars were also observed close in time to the science target. All observations were conducted in two bands simultaneously: a ``blue'' band centered at 562~nm with a width of 44~nm, and a ``red'' band centered at 832~nm with a width of 40~nm. The pixel scales of the ``blue'' and ``red'' EMCCDs are 0.0175649 arcsec/pixel and 0.0181887 arcsec/pixel, respectively. Reconstructed 256$\times$256 pixel images in each band were computed using the point source calibrator images following the approach by \citet{2011AJ....142...19H}. The background sensitivity of the reconstructed images was measured using a series of concentric annuli centered on the target star, resulting in 5$\sigma$ sensitivity limits (in $\Delta$-mags) as a function of angular separation (Fig.~\ref{fig:speckle_fig}). No secondary sources were detected in the reconstructed $\sim4.6\arcsec\times4.6\arcsec$ images.

\subsubsection{IRCS AO imaging}
\label{sec:AO}

High-resolution imaging was performed on 2017 May 22 for K2-180 and K2-140 by IRCS with the Subaru 8.2\,m telescope \citep{Kobayashi_et_al_2000} using the curvature AO system with 188 elements, AO188 \citep{Hayano_et_al_2010}. The high-resolution mode was selected at a pixel scale of 0.0206$\arcsec$ per pixel and the FoV of $21\arcsec\times21\arcsec$. Both targets were observed with the $H$-band filter and two different lengths of exposure times. The first sets were long-exposure frames with saturated stellar images in order to search for faint nearby sources around the target stars. The second set of exposures were unsaturated frames for the flux calibration. Both saturated and unsaturated frames were taken using five-point dithering with a dithering size of 2.5$\arcsec$. The total scientific exposure times for the saturated frames of K2-180 and K2-140 were 450\,s and 750\,s, respectively. The IRCS data were reduced to extract the median-combined, distortion-corrected images for saturated and unsaturated frames \citep{hirano_et_al_2016}. The full-width-at-half-maximum (FWHM) was measured for unsaturated images to be 0.114$\arcsec$ and 0.095$\arcsec$. A visual inspection revealed that no bright source is present in the FoV of K2-140, while two faint stars were identified 7.4$\arcsec$ northeast (NE) and 7.6$\arcsec$ southeast (SE) from K2-180. The two objects fall inside the photometric aperture (40$\arcsec$) used to extract the light curve of K2-180 and are thus sources of light contamination. 

\begin{figure}
\includegraphics[width=\columnwidth]{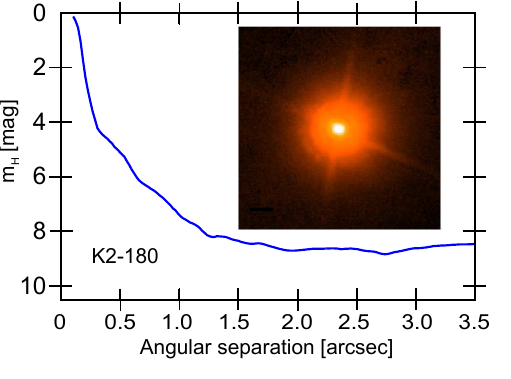}
\includegraphics[width=\columnwidth]{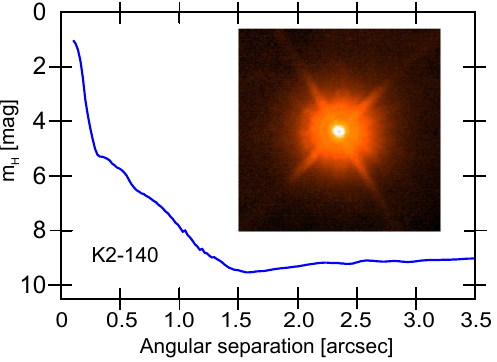}
\caption{$H$-band 5$\sigma$ contrast curves from the saturated images taken by Subaru/IRCS. \textit{Upper panel} K2-180. 
\textit{Lower panel}: K2-140. The insets display $4.0\arcsec\times4.0\arcsec$ images of each star.}
\label{fig:ircs}
\end{figure}

The two faint contaminants to K2-180 are listed in the SDSS12 catalog \citep{Alam_et_al_2015} and are identified as J082551.85+101451.8 and J082551.72+101441.1. Based on the SDSS $g$- and $r$-band magnitudes, the \kepler\ band magnitudes ($K_\mathrm{p}$) of both stars are estimated to be $K_\mathrm{p}\sim20$\,mag, which is consistent with a flux contrast of $\sim10^{-3}$ with respect to K2-180. The strong flux contrast implies that these faint objects cannot be the sources of the transit-like signals detected in the \ktwo\ time-series photometry of K2-180. Additionally, light curves were extracted using customized apertures that are centered around these faint stars and excluding a significant fraction of light from K2-180. The extracted light curves of the fainter nearby stars do not exhibit any deeper eclipses, indicating that K2-180 is the source of transits. The Subaru/IRCS's 5$\sigma$ contrast curves for each object are shown in Fig.~\ref{fig:ircs}.

\subsubsection{ALFOSC seeing-limited imaging}

\begin{figure}
\includegraphics[width=\columnwidth]{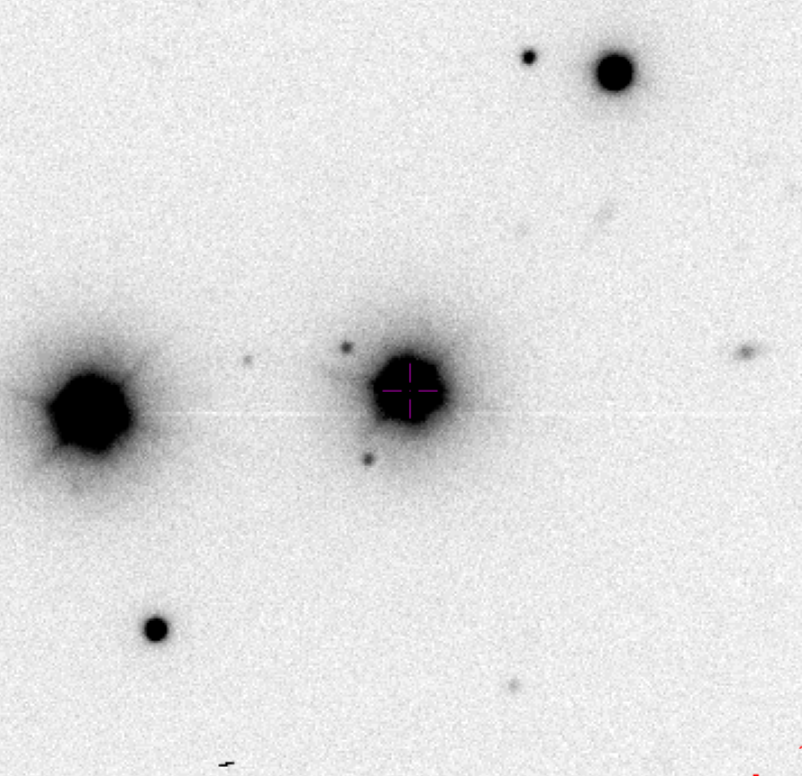}
\caption{ALFOSC $I$-band image of the region around K2-180. North is up, east is left. ALFOSC pixel scale is about $0.2\arcsec$ per pixel and the image covers a field of $1.25\arcmin\times1.25\arcmin$. The two nearby fainter companions are located at $\sim7\arcsec$ northeast and southeast of K2-180. Note that the Kepler pixel scale is 3.98$\arcsec$.}
\label{fig:alfosc}
\end{figure}

In order to measure a contamination factor arising from the two nearby stars, seeing-limited images of K2-180 were acquired with the ALFOSC camera mounted at the at the 2.56\,m Nordic Optical Telescope (NOT) of Roque de los Muchachos Observatory (La Palma, Spain). The observations were performed on 2017 January 10 as part of the observing program 56-209, setting the exposure time to 20\,s and using the Bessel $R$ and $I$ filters. ALFOSC has a FOV of $6.4\arcmin\times6.4\arcmin$ and a pixel scale of about $0.2\arcsec$/pixel. Fig.~\ref{fig:alfosc} shows the $1.25\arcmin\times1.25\arcmin$ portion of the $I$-band image centered around K2-180. The $I$-band and $R$-band magnitude differences between the two nearby stars and K2-180 are 7.15 and 7.44 for the contaminant to the NE of K2-180, and 7.18 and 8.00 for the contaminant to the SE, respectively. The magnitude of the two contaminants were placed into a color-density diagram \citep{Pecaut_mamajek_2013}. Under the assumption that they are main-sequence objects, these $\sim$K8V (NE) and $\sim$K1V (SE) contaminating stars are at $\sim$2000\,pc and $\sim$5700\,pc distance, while K2-180 is located at $\sim$210\,pc. Therefore they are not gravitationally bound to K2-180 but they form an optical triple. The two nearby stars produce a contamination of $0.2\pm0.1$\,\% that was taken into account while modeling the transit light curve. 

\subsubsection{High-resolution spectroscopy} 
\label{Sec:High_Res_Spec}

\begin{table*}
\centering
\caption{FIES and HARPS-N measurements of K2-180 and K2-140.}
\label{tab:RVs}
\begin{tabular}{lcccccccc}
\hline
\hline
\noalign{\smallskip}
$\mathrm{BJD_{TDB}}^{(\mathrm{a})}$ & RV & $\sigma_\mathrm{RV}$ & BIS & FWHM & log\,$R^\prime_\mathrm{HK}$ & $\sigma_\mathrm{log\,R^\prime_\mathrm{HK}}$ & $T_\mathrm{exp}$ & S/N         \\
 -2\,450\,000      & [\kms] &        [\kms]         &    [\kms]  & [\kms] &   &                     & [s]                & @5500~\AA   \\
\noalign{\smallskip}
\hline
\noalign{\smallskip}
\multicolumn{9}{c}{K2-180} \\
\noalign{\smallskip}
\hline
\noalign{\smallskip}
\multicolumn{4}{l}{FIES} \\
\noalign{\smallskip}
7833.417363 & -76.8499 & 0.0051 & $-$0.0396 & 11.5236 & - & - & 3600 & 35 \\
7834.487740 & -76.8549 & 0.0044 & $-$0.0337 & 11.5076 & - & - & 3600 & 42 \\
7835.507991 & -76.8546 & 0.0066 & $-$0.0371 & 11.5130 & - & - & 3600 & 29 \\
\noalign{\smallskip}
\multicolumn{4}{l}{HARPS-N} \\
\noalign{\smallskip}
 7692.757267 & -76.6127 & 0.0027 & -0.0462 &  6.1637 &  -4.955 &   0.037 &  2700 &  36.9 \\
 7693.742064 & -76.6208 & 0.0045 & -0.0339 &  6.1556 &  -4.995 &   0.083 &  2700 &  25.0 \\
 7762.768142$^\mathrm{(b)}$ & -76.6208 & 0.0056 & -0.0334 &  6.1435 &  -5.184 &   0.183 &  2880 &  22.3 \\
 7836.400872 & -76.6202 & 0.0036 & -0.0415 &  6.1489 &  -4.933 &   0.052 &  3600 &  31.1 \\
 7837.391740 & -76.6202 & 0.0034 & -0.0362 &  6.1651 &  -4.982 &   0.052 &  3600 &  32.9 \\
 7838.484832 & -76.6164 & 0.0026 & -0.0389 &  6.1635 &  -4.915 &   0.031 &  3600 &  40.3 \\
 7841.363546 & -76.6127 & 0.0032 & -0.0384 &  6.1695 &  -4.871 &   0.038 &  3600 &  34.8 \\
 7844.399520 & -76.6129 & 0.0032 & -0.0483 &  6.1733 &  -5.018 &   0.053 &  3600 &  34.7 \\
 7852.386932 & -76.6107 & 0.0036 & -0.0472 &  6.1720 &  -4.946 &   0.050 &  3600 &  31.0 \\
 7868.427489 & -76.6099 & 0.0026 & -0.0350 &  6.1589 &  -4.909 &   0.031 &  3600 &  40.4 \\
 7874.390843 & -76.6153 & 0.0023 & -0.0377 &  6.1664 &  -4.975 &   0.029 &  3600 &  44.1 \\
 7877.381999 & -76.6103 & 0.0022 & -0.0405 &  6.1506 &  -4.904 &   0.023 &  3300 &  44.7 \\
 8055.704014$^\mathrm{(b)}$ & -76.6296 & 0.0080 & -0.0337 &  6.1467 &  -4.969 &   0.166 &  3321 &  17.0 \\
 8076.763810 & -76.6195 & 0.0028 & -0.0316 &  6.1515 &  -4.947 &   0.040 &  3000 &  37.2 \\
\noalign{\smallskip}
\hline
\noalign{\smallskip}
\multicolumn{9}{c}{K2-140} \\
\noalign{\smallskip}
\hline
\noalign{\smallskip}
\multicolumn{4}{l}{FIES} \\
\noalign{\smallskip}
7833.576083 &  1.0567  & 0.0106 & -0.0224 & 11.5831 & - & - & 3600 & 34 \\
7834.612794 &  1.1230  & 0.0135 & -0.0299 & 11.6212 & - & - & 3600 & 29 \\
7835.559196 &  1.2080  & 0.0150 & -0.0216 & 11.5689 & - & - & 3600 & 25 \\
7836.653344 &  1.2290  & 0.0136 & -0.0354 & 11.6114 & - & - & 3000 & 27 \\
7845.574989 &  1.0472  & 0.0120 & -0.0058 & 11.6062 & - & - & 3000 & 33 \\
7865.495790 &  1.0241  & 0.0161 & -0.0418 & 11.6087 & - & - & 3000 & 21 \\
7867.483874 &  1.1422  & 0.0101 & -0.0366 & 11.5984 & - & - & 3000 & 32 \\
7877.463843 &  1.0787  & 0.0134 & -0.0437 & 11.5008 & - & - & 3000 & 35 \\
7890.421698 &  1.1436  & 0.0153 & -0.0184 & 11.5472 & - & - & 3000 & 25 \\
7893.444316 &  1.1165  & 0.0116 & -0.0110 & 11.5661 & - & - & 3600 & 35 \\
7894.470165 &  1.1840  & 0.0099 & -0.0336 & 11.5826 & - & - & 3600 & 37 \\
7895.411315 &  1.2368  & 0.0115 & -0.0243 & 11.6303 & - & - & 3600 & 35 \\
7896.512791 &  1.1859  & 0.0131 & -0.0232 & 11.6190 & - & - & 3000 & 28 \\
\noalign{\smallskip}
\hline
\hline
\end{tabular}
\begin{tablenotes}\footnotesize
  \item $^{(\mathrm{a})}$ Times are given in barycentric Julian date (BJD) in barycentric dynamical time (TDB). 
 \item $^{(\mathrm{b})}$ Affected by high airmass and bad sky conditions. Not included in our analysis.
\end{tablenotes}
\end{table*}

K2-140 and K2-180 were observed with FIES \citep{Frandsen1999,Telting2014} mounted at the NOT. Thirteen spectra of K2-140 and three spectra of K2-180 were collected between 2017 March 21 and May 23, as part of the observing programs P54-027 and P55-019. The high-resolution instrument set-up was used, which provides a resolving power of $R\approx67,000$ in the wavelength range of 3700--9100\,\AA. The exposure time was set to 2700--3600\,s according to sky conditions and time constraints of the observing schedule. Following the observing strategy as in \citet{Buchhave2010} and \citet{Gandolfi2013}, the intra-exposure RV drift of the instrument was traced by acquiring long-exposed ($T_\mathrm{exp}=35$\,s) ThAr spectra immediately before and after each observation. The data were reduced using standard IRAF and IDL routines, which include bias subtraction, flat fielding, order tracing and extraction, and wavelength calibration. Radial velocities were extracted via multi-order cross-correlations with the RV standard star HD\,50692 (G0V) and HD\,3765 \citep{Udry1999} for K2-140 and K2-180, respectively.

The RV follow-up of K2-180 was also performed by the HARPS-N spectrograph \citep[$R \approx 115,000$;][]{Cosentino2012} mounted at the 3.58~m Telescopio Nazionale Galileo (TNG) of Roque de los Muchachos Observatory (La Palma, Spain). Fourteen spectra were taken between 2016 October 31 and 2017 November 19, as part of the observing programs A34TAC\_10, A34TAC\_44, CAT16B\_61, OPT17A\_64, OPT17B\_59, and A36TAC\_12. The second fiber was used to monitor the sky background and the exposure time was set to 2700--3600~s. The data were reduced with the dedicated off-line HARPS-N pipeline and RVs were extracted by cross-correlating the extracted echelle spectra with a G2 numerical mask. 

The FIES and HARPS-N radial velocity measurements of K2-140 and K2-180 are listed in Table~\ref{tab:RVs}, along with their 1$\sigma$ uncertainties, the FWHM and bisector span (BIS) of the cross-correlation function (CCF), the Ca\,{\sc ii} H\,\&\,K activity index log\,$R^\prime_\mathrm{HK}$ (for the HARPS-N spectra only), the exposure time, and the S/N ratio per pixel at 5500\,\AA. K2-180 was observed at airmass higher than 2 on $\mathrm{BJD_{TDB}}=2457762.768142$ and under poor sky conditions on $\mathrm{BJD_{TDB}}=2458055.704014$, resulting in data with low S/N ratio. Both spectra were not included in the analysis.

Spectral line distortion caused by photospheric active regions (spots and plages) coupled to the stellar rotation and/or by blended eclipsing binary systems, induces an apparent RV variation. The lack of a significant correlation between RV and BIS (Table~\ref{tab:RVs}), as well as between RV and FWHM, can help to rule out false positives. The Pearson correlation coefficient between RV and BIS of K2-140 is 0.01 with a $p$-value of 0.99. The correlation coefficient for RV and FWHM is 0.27 with $p=0.36$. The coefficients for K2-180 are -0.36 with $p=0.21$, and 0.14 with $p=0.61$ for RV versus BIS and RV versus FWHM, respectively. Assuming a significance level of 0.05 for $p$ \citep{Fisher1925}, these quantities show no significant correlations. The periodograms of the BIS, FWHM, and log\,$R^\prime_\mathrm{HK}$ show no peaks with false-alarm probability lower than 20\,\%, indicating that the observed RV variation is very likely caused by the presence of the orbiting companions.

\section{Analysis}
\label{analysis}
\subsection{Stellar characterization}

In order to derive the fundamental parameters of the host stars (namely, mass $M_{\star}$, radius $R_{\star}$, and age), which are needed for a full interpretation of the planetary systems, the co-added FIES spectra of K2-140 (S/N\,$\sim$110) and the co-added HARPS-N spectra of K2-180 (S/N\,$\sim$120) were analyzed using the spectral analysis package Spectroscopy Made Easy (SME) \citep{Valenti_piskunov_1996,Valenti_fischer_2005,Piskunov_valenti_2017}. SME calculates synthetic stellar spectra for a set of given stellar parameters from grids of pre-calculated 1D/3D, LTE or non-LTE stellar atmosphere models. The code then fits the stellar models to the observed spectra of a given star using a least-squares procedure. By varying one or a few parameters and keeping others fixed, the true stellar parameters can be found with a good accuracy. The precision achievable is primarily dependent on the quality of the observed spectrum and the inherent precision of the utilized model grids. For K2-140 and K2-180, the non-LTE SME package version 5.2.2 together with the \texttt{ATLAS 12} model spectra grid \citep{Kurucz2013} were selected to fit the spectral features sensitive to the photospheric parameters.

The effective temperature, \teff, was determined from the profiles of the line wings of the H$_\mathrm{\alpha}$ and H$_\mathrm{\beta}$ \citep{Fuhrmann_et_al_1993,Fuhrmann_1994}. The cores of the lines were excluded because those originate from layers above the photosphere. The surface gravity \logg\ was estimated from the line wings of the Ca\,{\sc i}~$\lambda$6102, $\lambda$6122, $\lambda$6162 triplet, the Ca\,{\sc i} $\lambda$6439 line, and the Mg\,{\sc i} $\lambda$5167, $\lambda$5172, $\lambda$5183 triplet \citep{Fuhrmann_1997}. Many lines were simultaneously fit in different spectral regions to measure the metal abundances [Fe/H], [Ca/H] and [Mg/H]. The calibration equation for Sun-like stars from \citet{Bruntt2010b} was adopted to fix the microturbulent velocity \vmic. The projected stellar rotational velocity \vsini\ and the macroturbulent velocity \vmac\ were estimated by fitting the profile of several clean and unblended metal lines. The best-fitting model was checked with the Na doublet $\lambda$5889 and $\lambda$5896. 

The resulting effective temperatures and \logg\ of K2-180 and K2-140 are \teff\,=\,5110\,$\pm$\,107\,K and \logg\,=\,4.3\,$\pm$\,0.2\,dex, and 5585\,$\pm$\,120\,K and \logg\,=\,4.4\,$\pm$\,0.2\,dex, respectively. All values derived by SME are reported in Table~\ref{tab:radius_mass}. The spectral types of the host stars are then determined from the calibration scale for dwarf stars \citep{Pecaut_mamajek_2013} to be K2V and G6V, respectively. The interstellar extinction was estimated with the equation from \citet{Poznanski_et_al_2012} which uses the equivalent width of the Na absorption lines. This yielded to $A_\mathrm{V}=0$ for K2-180 based on the absence of interstellar components and to $A_\mathrm{V}=0.16^{+0.08}_{-0.05}$ for K2-140. The different distances (Table \ref{tab:stellartab2}) calculated with the \gaia\ parallaxes and with the absolute magnitudes corroborate also the estimated extinction values. Both distances agree well for K2-180 which implies that there is no extinction. For K2-140, however, both distances slightly disagree but are still consistent within 1$\sigma$ indicating a small extinction. Note that the small differences in the distances may also be due to a not accurate assumed $M_\mathrm{V}$ from the table of \citet{Pecaut_mamajek_2013}.

\begin{table*}
\centering
\caption{Stellar radii and mass determined by the different approaches for K2-180 and K2-140. Values in bold are adopted as the final stellar radii and masses.}
\label{tab:radius_mass}
\begin{tabular}{lcccccc}
\hline
\hline
\noalign{\smallskip}
source & stellar radius [\Rsun] & stellar mass [\Msun] & stellar density [\gcm] & \teff\ [K] & \logg\ [dex] & [Fe/H] [dex] \\
\noalign{\smallskip}
\hline
\noalign{\smallskip}
\multicolumn{7}{c}{K2-180} \\
\noalign{\smallskip}
\hline
\noalign{\smallskip}
SME&-&-&-&5110 $\pm$ 107&4.3 $\pm$ 0.2&-0.65 $\pm$ 0.10\\
\textbf{\texttt{BASTA}}$^{(\mathrm{a})}$&\textbf{0.69 $\pm$ 0.02}&\textbf{0.71 $\pm$ 0.03}&\textbf{3.04 $\pm$ 0.39}&5319 $\pm$ 55&4.6 $\pm$ 0.02&-0.5$^{+0.0}_{-0.2}$\\
\texttt{PARAM 1.3}$^{(\mathrm{a})}$ & 0.68 $\pm$ 0.02 & 0.72 $\pm$ 0.02 & 3.22 $\pm$ 0.37 & - & 4.6 $\pm$ 0.02 & -\\
\texttt{SpecMatch-Emp} & 0.82 $\pm$ 0.13 &  - & - & 5310 $\pm$ 110 & - & -0.47 $\pm$ 0.08\\
\gaia$^{(\mathrm{a})}$ & 0.79 $\pm$ 0.04 & -  & - & - &- &-\\
\gaia$^{(\mathrm{b})}$ & 0.69 $\pm$ 0.02 & - & - & - & - & -\\
\citet{torres_et_al_2010}$^{(\mathrm{a})}$  & 1.03 $\pm$ 0.27 & 0.78 $\pm$ 0.07 & 1.00 $\pm$ 0.88 & - & - & -\\
\citet{torres_et_al_2010}$^{(\mathrm{b})}$ & 1.06 $\pm$ 0.27 & 0.83 $\pm$ 0.08 & 0.98 $\pm$ 0.84 & - & - & -\\
\citet{Enoch_et_al_2010}$^{(\mathrm{a})}$ & 0.74 $\pm$ 0.07 & 0.83 $\pm$ 0.05 & 2.88 $\pm$ 0.99 & - & - & -\\
\citet{Enoch_et_al_2010}$^{(\mathrm{b})}$ & 0.75 $\pm$ 0.08 & 0.86 $\pm$ 0.05 & 2.87 $\pm$ 1.08 & - & - &-\\
\citet{Southworth_2011}$^{(\mathrm{a})}$ & 0.70 $\pm$ 0.07 & 0.63 $\pm$ 0.06 & 2.58 $\pm$ 1.02 & - &- &-\\
\citet{Southworth_2011}$^{(\mathrm{b})}$ & 0.71 $\pm$ 0.07 & 0.65 $\pm$ 0.06 & 2.55 $\pm$ 0.99 & - &- &-\\
\noalign{\smallskip}
\hline
\noalign{\smallskip}
\multicolumn{7}{c}{K2-140} \\
\noalign{\smallskip}
\hline
\noalign{\smallskip}
{SME} & - & - & - & 5585 $\pm$ 120 & 4.4 $\pm$ 0.2 & 0.10 $\pm$ 0.10\\
\textbf{\texttt{BASTA}}$^{(\mathrm{a})}$ & \textbf{1.06$^{+0.07}_{-0.06}$} & \textbf{0.96$^{+0.06}_{-0.04}$} & \textbf{1.13 $\pm$ 0.28} & 5694$^{+83}_{-76}$ & 4.4$^{+0.07}_{-0.06}$ & 0.1$^{+0.08}_{-0.04}$\\
\texttt{PARAM 1.3}$^{(\mathrm{a})}$ & 1.01 $\pm$ 0.05 & 0.98 $\pm$ 0.05 & 1.33 $\pm$ 0.27 & - & 4.40 $\pm$ 0.05 & -   \\
\texttt{SpecMatch-Emp} & 1.00 $\pm$ 0.16 & - & - & 5711 $\pm$ 110 & - & 0.24 $\pm$ 0.08 \\
\gaia$^{(\mathrm{a})}$ & 1.13 $\pm$ 0.08 & -  & - & - &- &-  \\
\gaia$^{(\mathrm{b})}$ & 1.07 $\pm$ 0.08 & -  & - & - &- &-  \\
\noalign{\smallskip}
\hline
\hline
\end{tabular}
\begin{tablenotes}\footnotesize
  \item $^{(\mathrm{a})}$ Calculated with \teff\,=\,5110\,$\pm$\,107\,K for K2-180 and 5585\,$\pm$\,120\,K for K2-140 from SME. 
 \item $^{(\mathrm{b})}$ Calculated with \teff\,=\,5310\,$\pm$\,110\,K from \texttt{SpecMatch-Emp} for K2-180 and \teff\,=\,5711\,$\pm$\,110\,K for K2-140.
\end{tablenotes}
\end{table*}

Stellar masses, radii, and ages of the two stars were determined using the BAyesian STellar Algorithm (\texttt{BASTA}) \citep{Aguirre_et_al_2015} with a grid of the Bag for Stellar Tracks and Isochrones (BaSTI) isochrones \citep{Pietrinferni_et_al_2004}. The spectroscopic parameters \teff, \logg\ and [Fe/H] from SME (Table~\ref{tab:radius_mass}), the spectral energy distribution (SED) using BVJHKgri-band photometry (Table~\ref{tab:stellartab}), and the \gaia\ Data Release 2 (DR2) parallaxes (Table~\ref{tab:stellartab2}) were used as input for the modeling. \texttt{BASTA} derives stellar parameters in a Bayesian scheme by simultaneously fitting all observables to a finely-sampled grid of precomputed stellar isochrones. The ($16\%$, $50\%$, $84\%$) quantiles of the posteriors derived by \texttt{BASTA} are reported. Apparent magnitudes are converted to absolute magnitudes using the exponentially decreasing space density (EDSD) prior on the parallax \citep{Astraatmadja_Bailer_Jones_2016} taking into account the estimated absorption in each bandpass. A conservative systematic uncertainty of $1\%$ was added to the apparent magnitudes to account for any potential systematics between filter systems. \texttt{BASTA} estimated a stellar mass and radius of $M_{\star}=0.71\pm0.03$\,\Msun\ and $R_{\star}=0.69\pm0.02$\,\Rsun\ for K2-180, and of $M_{\star}=0.96^{+0.06}_{-0.04}$\,\Msun\ and $R_{\star}=1.06^{+0.07}_{-0.06}$\,\Rsun\ for K2-140. The system K2-180 has an age of 9.5$^{+4.0}_{-5.6}$ Gyr and the system K2-140 is 9.8$^{+3.4}_{-4.6}$ Gyr old. The uncertainties derived by \texttt{BASTA} are internal to the BaSTI isochrones and do not include systematics related to the choice of input physics. It is worth knowing that using \texttt{BASTA}, \teff, \logg, and [Fe/H] for K2-180 are relatively poor fit since the isochrones prefer a larger \logg\ of 4.6\,$\pm$\,0.2\,dex and a hotter \teff\ of 5319\,$\pm$\,55\,K compared to the values from SME (Table~\ref{tab:radius_mass}), whereas all value agree with SME for K2-140.

For an independent check on the \texttt{BASTA} results, stellar masses and radii were also derived with different methods: \texttt{PARAM 1.3}\footnote{\url{http://stev.oapd.inaf.it/cgi-bin/param_1.3}} \citep{daSilva_et_al_2006}, \texttt{SpecMatch-Emp} \citep{Yee_et_al_2017} and combining the \gaia\ distance with \teff. All values for the stellar radii, masses and densities determined by the different approaches as well as other estimated quantities (\teff, \logg\ and [Fe/H]) are summarized in Table~\ref{tab:radius_mass}.

The Bayesian \texttt{PARAM 1.3} online applet was used with the \texttt{PARSEC} isochrones from \citet{bressan_et_al_2012}. This tool needs \teff, [Fe/H], parallax, and the apparent visual magnitude. The code also estimates the \logg\ which is for K2-180 slightly larger just as the \logg\ derived by \texttt{BASTA}.

\texttt{SpecMatch-Emp} relies on empirical spectra and compares the observed spectra to a library of well-characterized stars (M5 to F1) observed by Keck/HIRES.\texttt{SpecMatch-Emp} also calculates \teff\ and \logg\ which agree within 1$\sigma$ with the values derived by SME. The higher \teff\ of 5310 $\pm$ 110 K for K2-180 is also in agreement with the preferred higher temperature from \texttt{BASTA}.

The calculation of the stellar radii combining the \gaia\ distance with \teff\, and the apparent visual magnitude without the use of isochrones or libraries assumes $A_\mathrm{V}=0$ for K2-180 and $A_\mathrm{V}=0.16^{+0.08}_{-0.05}$ for K2-140 and the bolometric correction from \citet{torres_2010}.

For K2-180, the stellar radius derived by the different approaches agrees only within 2$\sigma$. To further check on this discrepancy, stellar radii and masses were also estimated using the calibration equations from \citet{torres_et_al_2010}, \citet{Enoch_et_al_2010} and \citet{Southworth_2011}. The \citet{torres_et_al_2010} equations need \teff, \logg, and [Fe/H] as input values and were calibrated with 95 eclipsing binaries where the masses and radii are known to be better than 3$\%$. The advantage of the other calibration equations from \citet{Enoch_et_al_2010} and \citet{Southworth_2011} is that the input is \teff, [Fe/H], and the density which is derived from the transit modeling. \citet{Enoch_et_al_2010} calibrated their equations with a subsample out of the 95 eclipsing binaries from \citet{torres_et_al_2010} with measured metallicities. The database from \citet{Southworth_2011} consisted of 90 detached eclipsing binaries with masses up to 3 solar masses and measured metallicities. 

The values calculated with the \citet{torres_et_al_2010} equations are completely off when comparing the spectroscopic derived stellar density ($\rho_{\star}\sim$\,1.00\,$\pm$\,0.88\,\gcm) with the density derived from the LC+RV fit ($\rho_{\star}$\,=\,2.63\,$\pm$\,0.67\,\gcm) and should therefore not be trusted. One reason for this could be that the \citet{torres_et_al_2010} equations need \logg\ as input which is only weak constrained using SME measured from the line wings.The values calculated by the equations from \citet{Enoch_et_al_2010} and \citet{Southworth_2011} show no significant difference depending on the \teff.

The values derived by \texttt{BASTA} are taken as the final values for the stellar radius, mass and age because of two reasons. First, since K2-180 is a metal-poor star, the \logg\ is hard to measure from the spectral line wings. Second, the higher \texttt{SpecMatch-Emp} temperature is preferred by \texttt{BASTA} and the radii calculated using the \gaia\ distance, \teff\ and the apparent visual magnitude. Because the different \teff\ agree within 1$\sigma$ and the true \teff\ may be somewhere between the value calculated by \texttt{SpecMatch-Emp} and SME, the values estimated by SME are reported together with the final adopted stellar parameters for K2-180 and summarized in Table~\ref{tab:stellartab2}.

Stellar radii derived for K2-140 agree within 1$\sigma$ for all different approaches (see Table~\ref{tab:radius_mass}). Therefore, it is justified to take the results from \texttt{BASTA} as the final values for the stellar radii, masses, and ages for K2-140. Using the spectroscopic derived stellar density of 1.13 $\pm$ 0.28 \gcm\ estimated from the parameters derived by \texttt{BASTA}, the expected value for $a/R_\star$ is 13.7 $\pm$ 1.1 which is in good agreement with the one derived from the LC+RV fit (see Sec.~\ref{sec:joint}). The final adopted stellar parameters for K2-140 are summarized in Table~\ref{tab:stellartab2}.

The rotation period \Prot\ of a star can be measured from the quasi-periodic photometric variability induced by the presence of active regions carried around by stellar rotation. The \ktwo\ light curve of K2-180 shows no significant quasi-periodic flux variation (Fig.~\ref{fig:C5_9617_detection}). Although the light curve of K2-140 shows instead photometric variability (Fig.~\ref{fig:C10_5255_detection}), the data gap combined with the relatively short baseline hampers a reliable derivation of \Prot. Therefore, the stellar rotational periods were estimated using the projected rotational velocity \vsini\ combined with the stellar radius, under the assumption that both stars are seen equator-on. The stellar rotation period of K2-140 and K2-180 were found to be \Prot$=14.6\pm4.1$\,days and \Prot$=15.7\pm7.5$\,days, respectively.

\begin{table*}
\centering
 \caption{Stellar parameters of K2-180 and K2-140 adopted in this paper.} 
 \label{tab:stellartab2}  
 \begin{tabular}{lcc}
 \hline
 \hline
 \noalign{\smallskip}
  Parameter & K2-180 & K2-140 \\
  \noalign{\smallskip}
  \hline
  \noalign{\smallskip}
  Effective temperature \teff\ [K]  & 5110 $\pm$ 107 & 5585 $\pm$ 120 \\
  Surface gravity \logg\ [dex] & 4.3 $\pm$ 0.2  & 4.4 $\pm$ 0.2\\
  $\mathrm{[Fe/H]}$ [dex] & -0.65 $\pm$ 0.10 & +0.10 $\pm$ 0.10\\
  $\mathrm{[Ni/H]}$ [dex] & -0.70 $\pm$ 0.10 & +0.20 $\pm$ 0.10\\
  $\mathrm{[Ca/H]}$ [dex] & -0.45 $\pm$ 0.10 & +0.12 $\pm$ 0.10\\
  $\mathrm{[Mg/H]}$ [dex] & - & +0.27 $\pm$ 0.1\\
  $\mathrm{[Na/H]}$ [dex] & - & +0.12 $\pm$ 0.1\\
  Microturbulent velocity \vmic\ [\kms]  & 0.8 $\pm$ 0.3 & 1.03 $\pm$ 0.3\\
  Macroturbulent velocity \vmac\ [\kms] & 1.8 $\pm$ 1 & 1.5 $\pm$ 1\\
  Rotational velocity \vsini\ [\kms] & 2.1 $\pm$ 1.0 & 3.6 $\pm$ 1.0\\
 Spectral type & K2V & G6V \\
 Stellar mass $M_\star$ [\Msun]                           & 0.71 $\pm$ 0.03   & 0.96$^{+0.06}_{-0.04}$\\
 Stellar radius $R_\star$ [\Rsun]                         & 0.69 $\pm$ 0.02   & 1.06$^{+0.07}_{-0.06}$\\
  $\rho_{\star}$ [\gcm]$^{(\mathrm{a})}$                       & 2.63 $\pm$ 0.67        & 1.23 $\pm$ 0.05          \\
$\rho_{\star}$ [\gcm]$^{(\mathrm{b})}$	&	3.04 $\pm$ 0.39	&	1.13 $\pm$ 0.28	\\
  Stellar age [Gyrs]                                       & 9.5$^{+4.0}_{-5.6}$  & 9.8$^{+3.4}_{-4.6}$\\  
  Stellar rotation period \Prot\ [days] &  15.7 $\pm$ 7.5  &  14.6 $\pm$ 4.1 \\
  Distance $d$ [pc]$^{(\mathrm{d})}$				 & 206 $\pm$ 37	& 318 $\pm$ 26 \\
    Distance $d$ [pc]$^{(\mathrm{c})}$				 & 205 $\pm$ 5	& 351 $\pm$ 15 \\
 \noalign{\smallskip}
  \hline
  \hline
 \end{tabular}
 \begin{tablenotes}\footnotesize
 \item $^{(\mathrm{a})}$ Calculated from period and masses via Kepler's third law during the transit fit, not from RV.
 \item $^{(\mathrm{b})}$ Calculated from stellar radius and stellar mass.
 \item $^{(\mathrm{c})}$ Calculated form \gaia\ parallax\footnote{\url{https://gea.esac.esa.int/archive}}\,\citep{gaia_mission_2016,brown_et_al_gaia_2018}. Note that for the parallax error 0.1 mas was added quadratically to the parallax uncertainties to account for systematic errors of \gaia's astrometry \citep{Luri_et_al_2018}.  
 \item $^{(\mathrm{d})}$ Calculated with the absolute magnitudes that are determined from the calibration scale for dwarf stars from \citet{Pecaut_mamajek_2013} and assuming $A_\mathrm{v}=0$.
 \end{tablenotes}
\end{table*}

Assuming equatorial coordinates and proper motion from Table~\ref{tab:stellartab}, distances calculated with \gaia\ parallax from Table~\ref{tab:stellartab2}, and the systemic velocity from Table~\ref{tab:planparams_c5_9617}, the heliocentric space velocities are calculated. Following \citet{Ramirez_et_al_2007}, the probabilities of the population membership are calculated (Table~\ref{tab:prob}). It is therefore most likely that K2-180 belongs to the thick disc population and K2-140 to the thin disc population. This conclusion also agrees with the spectroscopically measured [Fe/H] values.

\begin{table}
\centering
 \caption{Population membership probabilities after \citet{Ramirez_et_al_2007}.} 
 \label{tab:prob}  
 \begin{tabular}{lcc}
 \hline
 \hline
 \noalign{\smallskip}
 membership & K2-180 & K2-140 \\
  \noalign{\smallskip}
  \hline
  \noalign{\smallskip}
  thin  & 0.23 $\pm$ 0.07 & 0.99 $\pm$ 0.07 \\
  thick & 0.73 $\pm$ 0.06  & 0.0075 $\pm$ 0.0008\\
  halo & 0.033 $\pm$ 0.001 & 0.000050 $\pm$ 0.000003\\
 \noalign{\smallskip}
  \hline
  \hline
 \end{tabular}
\end{table}

\subsection{Combined RV and light curve modeling}
\label{sec:joint}

A combined analysis of the \ktwo\ stellar light curves from \citet{vanderburg_johnson_2014} and the RV data for each system was performed using the Transit and Light Curve Modeller Code (\texttt{TLCM}), as done in previous KESPRINT publications \citep[e.g., in][]{Smith_et_al_2018}. The software code is described in detail in \citet{csizmadia_et_al_2011,csizmadia_et_al_2015} and Csizmadia (2018, under revision). \texttt{TLCM} models both the light curve and RV measurements simultaneously. In calculating the transit curve \texttt{TLCM} uses the quadratic limb-darkening model from \citet{mandel_agol_2002}. The program can calculate eccentric orbits with the inclusion of an overall RV drift. The fit is performed by minimizing
\begin{equation} \label{eq:x2}
\chi^2 = \frac{1}{N_\mathrm{LC}}\sum\limits_{i=1}^{N_\mathrm{LC}}\left( \frac{f_\mathrm{i} - f_\mathrm{m,i}}{\sigma_\mathrm{LC,i}} \right)^2 + \frac{1}{N_\mathrm{RV}} \sum\limits_{j=1}^{N_\mathrm{RV}} \left( \frac{RV_\mathrm{j} - RV_\mathrm{m,j}}{\sigma_\mathrm{RV,j}} \right)^2,
\end{equation}
where $N_\mathrm{LC}$ and $N_\mathrm{RV}$ are the total number of photometric points and RV points that were used in the fit. The quantities $f_\mathrm{i}$, $RV_\mathrm{j}$ and $f_\mathrm{m,i}$, $RV_\mathrm{m,j}$ are the observed and simulated photometric and RV points, respectively. The uncertainties $\sigma_\mathrm{LC,i}$ and $\sigma_\mathrm{RV,j}$ refer to the photometric and RV measurements. The $\chi^2$ values were simply the sum of the individual $\chi^2$. The fit is optimized by a Genetic Algorithm approach \citep{Geem_2001} in order to find a good starting point for the following Simulated Annealing analysis \citep{kallrath_milone_2009}. The Simulated Annealing is similar to the Markov Chain Monte Carlo (MCMC) analysis except that the acceptance probability is continuously decreased during the optimization process. The results of the Simulated Annealing and bootstrap-analysis are used to refine the results obtained by Genetic Algorithm and to estimate the 1$\sigma$ error. 

Prior to the analysis, segments twice as long as the transit duration and centered around each transit were extracted from the \ktwo\ light curves of K2-180 and K2-140. Parabolic functions were fit to these out-of-transit points. Each segment is divided by the corresponding parabola and the light curve was folded at the detected orbital period of the planets. A total of 166 and 289 photometric data points with exposure times of $\sim$30 minutes each were eventually extracted from the light curves of K2-180 and K2-140, respectively. 

The fit parameters for the combined LC+RV fit are the orbital period, the epoch, the scaled semi-major axis $a/R_\star$, the planet-to-star radius ratio $R_\mathrm{p}/R_\star$, the impact parameter $b$, the limb-darkening coefficient combinations $u_\mathrm{+} = u_\mathrm{a} + u_\mathrm{b}$ and $u_\mathrm{-} = u_\mathrm{a} - u_\mathrm{b}$, where $u_\mathrm{a}$ and $u_\mathrm{b}$ are the linear and quadratic limb darkening coefficients. Further fit parameters are the flux-shift which is able to correct possible normalization errors and the third light \citep{csizmadia_et_al_2013} within prescribed limits (0 for K2-140b and 0.2 $\pm$ 0.1 for K2-180b) to take contamination into account. The parameterization of the eccentricity and the argument of pericenter $e \cos \omega$ and $e \sin \omega$, the radial velocity amplitude $K$ of the star, the systematic velocity $V_\mathrm{\gamma}$ and the RV-offsets of the different spectrographs are also fit.

The Bayesian Information Criterion (BIC) \citep{Kass_Raftery_1995} is used to determine if a circular or eccentric orbit solution is favored. BIC is a better choice than $\chi^2$ for an acceptable final solution because it takes the number of free parameters into account, which may vary from case to case.

Another independent analysis was run with the MCMC code \texttt{pyaneti} \citep{Barragan_et_al_2018}. The MCMC solution is consistent within the 1$\sigma$ confidence level for K2-180b and within 3$\sigma$ for K2-140b. The latter discrepancy is explained in Sec. \ref{sec:k2_140}.  

\subsubsection{K2-180b}
\label{sec:k2_180}
\begin{figure}
\includegraphics[width=\columnwidth]{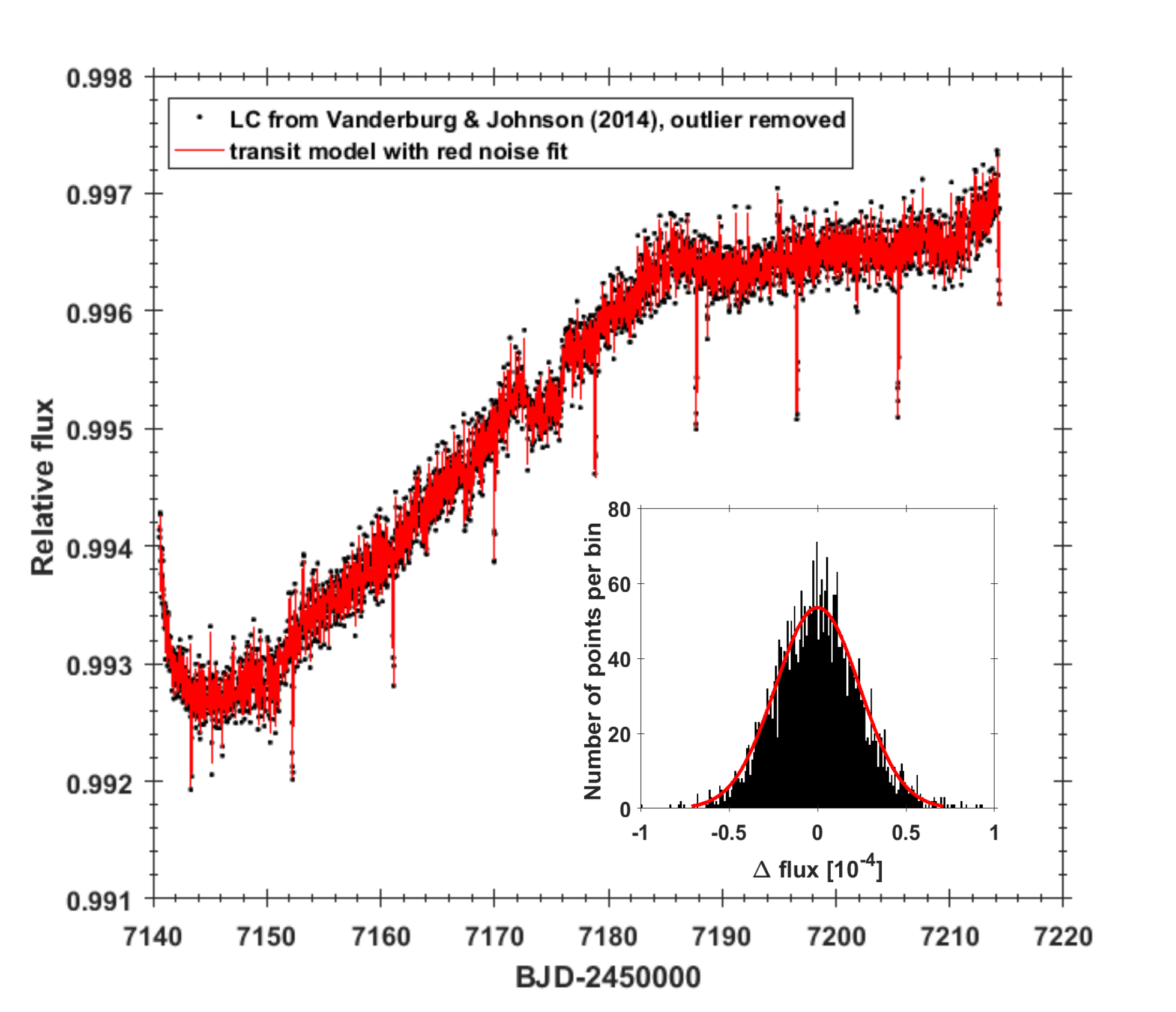}
\caption{Light curve of K2-180b from \citet{vanderburg_johnson_2014} after a 3$\sigma$-clip is applied.
The best-fit red noise model is shown in red. The inset displays the histogram of the residuals between the LC from \citet{vanderburg_johnson_2014} after the 3$\sigma$-clip and the red noise+transit model fit. The black bars in the inset shows the number of the residual points in the bin and the red curve is a Gaussian fit to the histogram.}
\label{fig:C5_9617_rednoise}
\end{figure}

The modeling of the RV and LC data from K2-180b included a red noise component (Fig.~\ref{fig:C5_9617_rednoise}) because the LC still showed large variations after extracting the segments centered around each transit. The red noise was modeled after the wavelet formulation from \citet{carter_winn_2009}. The impact parameter was set to $0<b<1$ to avoid over-fitting the data. 

A $3\sigma$-clip was also applied to the outliers relative to the transit model+red noise model fit because the LC from \citet{vanderburg_johnson_2014} has outliers which affect the subsequent analysis. The LC is likely affected by either instrumental systematics rather than by spot-crossing events because K2-180 is an inactive star. A poor decorrelation of the photometric flux with the pixel position in the \citet{vanderburg_johnson_2014} data could also have affected the quality of the LC. The histogram of the residuals (inset of Fig.~\ref{fig:C5_9617_rednoise}) shows a 47~ppm/30 min standard deviation. This, scaled to 6 hours, is 13ppm/$\sqrt{6}$ which is comparable to the value found by \citet{vanderburg_johnson_2014} for this magnitude range. 

\begin{figure}
\includegraphics[width=\columnwidth]{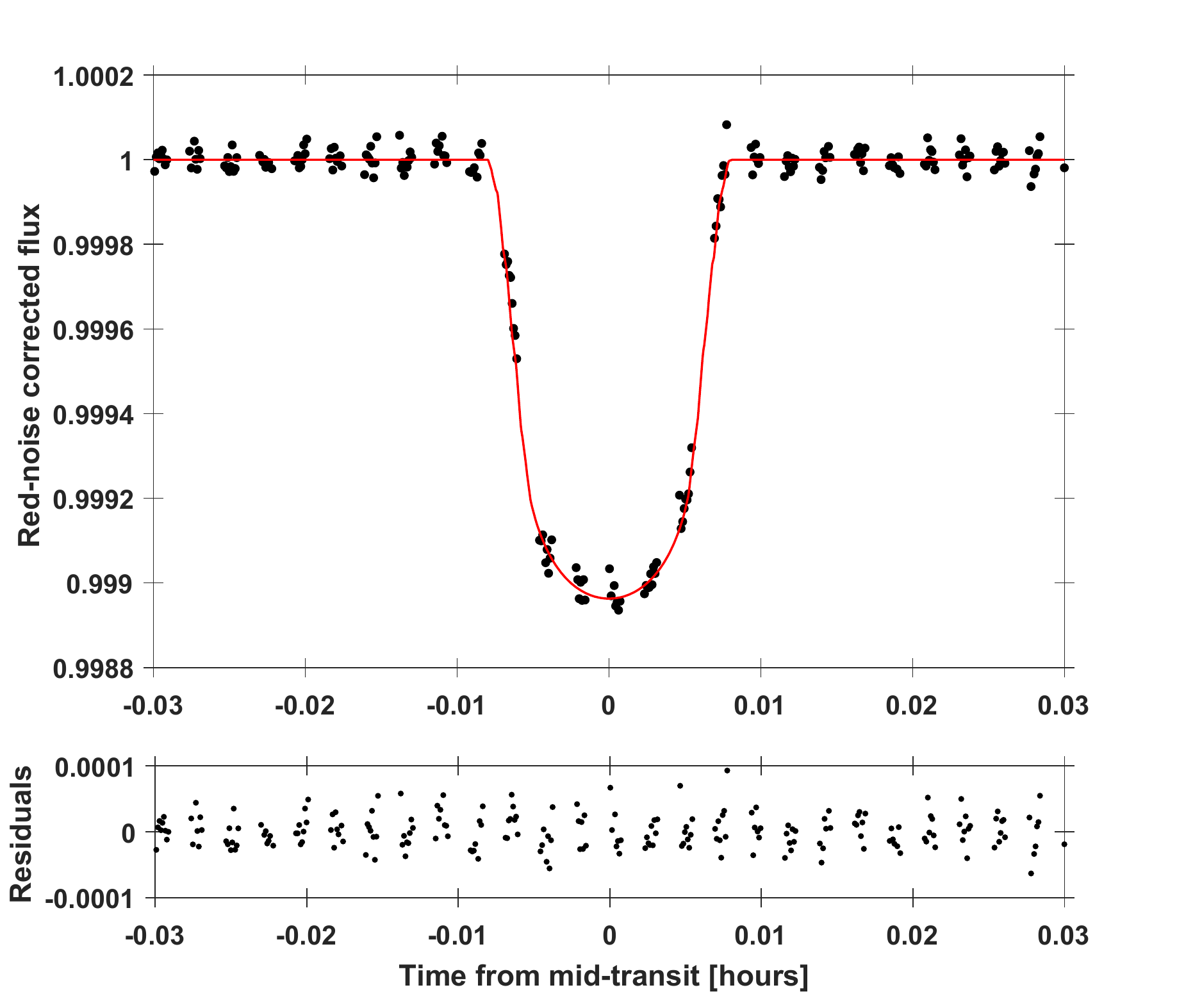}
\caption{\textit{Upper panel}: Best transit model for K2-180b (red solid line). Filled black circles are the folded light curve
corrected for red-noise effects that include stellar variability and instrumental noise. Note the visible clustering is produced by the red noise correction. \textit{Lower panel}: Fit residuals.}
\label{fig:c5_9617_transit}
\end{figure}

\begin{figure}
\includegraphics[width=\columnwidth]{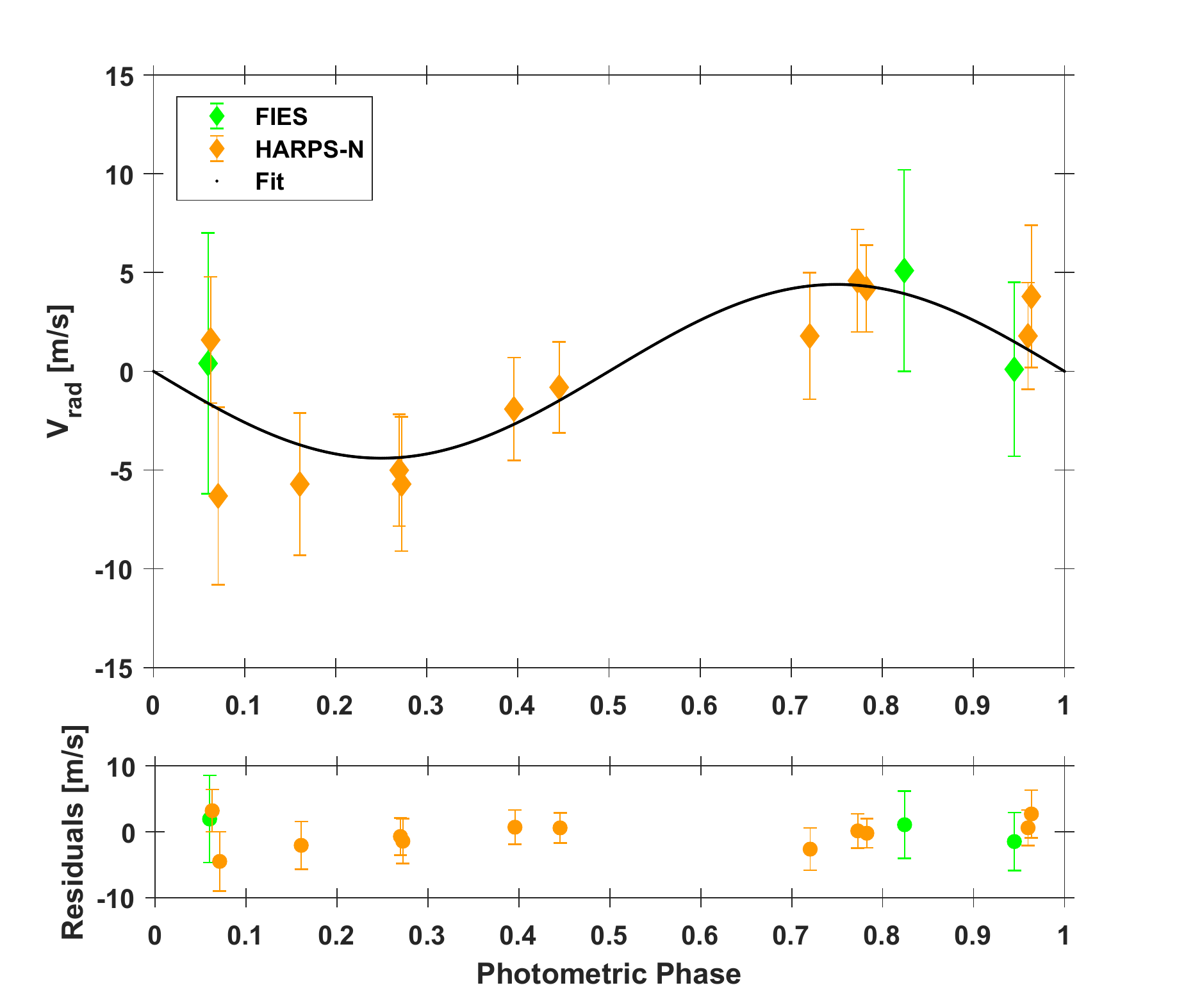}
\caption{\textit{Upper panel}: The best fit to the radial velocity values of K2-180b (black solid line). Green diamonds are the FIES measurements and the orange diamonds are the HARPS-N values. \textit{Lower panel}: Fit residuals.}
\label{fig:c5_9617_rv}
\end{figure}

The transit fit and RV fit are shown in Fig.~\ref{fig:c5_9617_transit} and Fig.~\ref{fig:c5_9617_rv}, respectively for K2-180b. The derived parameters are listed in Table~\ref{tab:planparams_c5_9617}.  

A negative jitter value was calculated because the  $\chi^{2}_{\mathrm{RV}}/N_{\mathrm{RV}}$ is around 0.5. This may indicate that the RV errors are probably overestimated according to e.g. \citet{Baluev_2009}. However, this can be merely just a statistical fluctuation of the $\chi^2$ distribution. Therefore, the following numerical experiment was performed: 1 million synthetic RV data-sets were produced with normally distributed random noise for 15 data points representing the number of RV observations for K2-180. The corresponding $\chi^{2}_{\mathrm{RV}}/N_{\mathrm{RV}}$ value was calculated and it turned out that in less than 5$\%$ of the cases the $\chi^{2}_{\mathrm{RV}}/N_{\mathrm{RV}}$ is between 0.4 and 0.6. This means that there is a non-negligible chance that the observed lower $\chi^2$ values of the RV-curve is due to a random residual-distribution effect rather than an overestimation of the errors. Therefore no RV jitter was included in the combined fit.  

\begin{table*}
\centering
\caption{Final adopted physical and geometrical parameters of the K2-180 and K2-140 systems. The convention, $\omega=90\degr$, for circular orbits is used so that $T_\mathrm{p}=T_\mathrm{0}$.} 
\label{tab:planparams_c5_9617}
\begin{tabular}{lrr}
\hline
\hline
\noalign{\smallskip}
  \, & K2-180 & K2-140 \\
\noalign{\smallskip}
\hline 
\noalign{\smallskip}
\multicolumn{1}{l}{Determined from photometry} \\
\noalign{\smallskip}
Epoch of transit $T_{0}$ [BJD$-$2450000]	& 7143.390 $\pm$ 0.002 & 7588.28509 $\pm$ 0.00005\\
Orbital period [days]						& 8.8665 $\pm$ 0.0003	& 6.569188 $\pm$ 0.000031$^{(\mathrm{a})}$ \\
Duration of the transit [hours]             & 2.98 $\pm$ 0.07       & 3.81 $\pm$ 0.04 \\
Depth of the transit [\%]                   & 0.12 $\pm$ 0.05	   & 1.557 $\pm$ 0.002\\
\noalign{\smallskip}
\hline
\noalign{\smallskip}
\multicolumn{1}{l}{Determined from combined LC + RV fit} \\
\noalign{\smallskip}
Orbital eccentricity $e^{(\mathrm{a})}$                      & 0                   & 0 \\
Argument of periastron $\omega$ [deg]$^{(\mathrm{a})}$       & 90                  & 90\\
RV semi-amplitude $K$ [\ms]							& 4.4 $\pm$ 0.7		  & 104.1 $\pm$ 2.7 \\
Systemic velocity  $V_{\gamma}$ [\kms]              & -76.855 $\pm$ 0.001   & 1.215 $\pm$ 0.004 \\
RV offset HARPS-N-FIES [\ms]                        & 240.5 $\pm$ 1.1       & -\\
RV offset HARPS-CORALIE [\ms]                       & -                   & 32.8 $\pm$ 5.2    \\
RV offset FIES-CORALIE  [\ms]                       & -                   & -1139.7 $\pm$ 4.8   \\
$a/R_\star$                                         & 22.2 $\pm$ 1.9        & 14.1 $\pm$ 0.2       \\
$R_\mathrm{p}/R_\star$                              & 0.0297 $\pm$ 0.0008    & 0.117 $\pm$ 0.001    \\
$b$                                                 & 0.4 $\pm$ 0.2         & 0.42 $\pm$ 0.04        \\
$i_\mathrm{p}$ [deg]                                & 88.9 $\pm$ 0.7        & 88.3 $\pm$ 0.1       \\
${u_+}$                                       & 0.7$^{(\mathrm{a})}$                 & 0.42 $\pm$ 0.09       \\
${u_-}$                                       & 0.1$^{(\mathrm{a})}$                 & 0.68 $\pm$ 0.27        \\
contamination [\%]                                  & 0.2 $\pm$ 0.1         & -          \\
flux-shift [ppm]                  & -4571 $\pm$ 7         & 0.15 $\pm$ 0.01              \\
$\sigma_{red}$ flux [ppm]               & 7640 $\pm$ 40         & -              \\
$\sigma_{white}$ flux [ppm] 	        & 60 $\pm$ 10           & -\\ 
\noalign{\smallskip}
\hline
\noalign{\smallskip}
\multicolumn{1}{l}{Absolute dimensions of the system} \\
\noalign{\smallskip}
Orbital semi-major axis $a$ [AU]$^{(\mathrm{b})}$              & 0.075 $\pm$ 0.001          & 0.068 $\pm$ 0.001 \\
Planetary mass   $M_\mathrm{p}$ [\Mjup]               & 0.036 $\pm$ 0.006 & 0.93 $\pm$ 0.04\\
Planetary radius $R_\mathrm{p}$ [\Rjup]               & 0.200 $\pm$ 0.011   & 1.21 $\pm$ 0.09\\
Planetary mean density $\rho_\mathrm{p}$ [g\,cm$^{-3}$] & 5.6 $\pm$ 1.9  & 0.66 $\pm$ 0.18\\
Equilibrium temperature $T_\mathrm{eq}$	[K]$^{(\mathrm{c})}$				 & 729 $\pm$ 49	& 962 $\pm$ 28 \\
Insolation flux $F$	[F$_{\earth}$]				 & 67 $\pm$ 14	& 204 $\pm$ 10 \\
\noalign{\smallskip}
\hline
\hline
\end{tabular}
\begin{tablenotes}\footnotesize
\item Note: Values are calculated assuming solar, Jupiter and Earth radii and masses of 1.98844 $\cdot10^{30}$ kg and 696,342 km, 1.89813 $\cdot10^{27}$ kg and 71,492 km and 5.9722$\cdot10^{24}$ kg and 6,378 km, respectively.
  \item  $^{(\mathrm{a})}$ Fixed parameter.
   \item $^{(\mathrm{b})}$ From modeling results.
   \item $^{(\mathrm{c})}$ Calculated with equation 3 from \citet{batalha_et_al_2013} with f=1 and Bond albedo of 0.3.
\end{tablenotes}
\end{table*}

\subsubsection{K2-140b}
\label{sec:k2_140}
The combined LC+RV fit of K2-140b includes the LC from \citet{vanderburg_johnson_2014}, the RV measurements presented in this paper, and the Doppler measurements reported in G18. G18 found an eccentric orbital solution with $e=0.120^{+0.056}_{-0.046}$ (2.6$\sigma$ significance). To further investigate the non-zero eccentricity found by G18, six different analysis of the two RV datasets were performed: 
\begin{enumerate}
\item  RV data from G18 with an elliptical orbit
\item RV data from G18 with a circular orbit
\item RV data by KESPRINT (K) with an elliptical orbit
\item RV data by KESPRINT with a circular orbit
\item Combined (Comb) RV datasets with an elliptic orbit
\item Combined RV datasets with a circular orbit
\end{enumerate}

For the specific case (i), the eccentricity is $e=0.08 \pm 0.03$ which is lower than the value of $e=0.120^{+0.056}_{-0.046}$ found by G18 but within their error. A fit on the same data sets was also performed with the software RVLIN \citep{wright_howard_2009}, fixing the orbital period and using the transit time as a constraint. The results ($e=0.084$, $\omega = 147.5$\degr, $K=113.5$\,\ms\ and $V_\gamma=1.2140$\,\kms) are in perfect agreement with our solution ($e=0.084\pm0.03$, $\omega = 144\pm27$\degr, $K=113\pm4$\,\ms\ and $V_\gamma=1.215\pm0.006$\,\kms). Since G18 gave few details about the fitting procedure (number of chains, chain length, convergence check, and stopping criterion) it is not possible to discuss discrepancies between the two results. The fits with $e=0$ result in smaller BIC values than the fits with $e\neq0$ with differences up to 11 ($\Delta\mathrm{BIC}_\mathrm{G18}=10.7$,~$\Delta\mathrm{BIC}_\mathrm{K}=11.0$ and $\Delta\mathrm{BIC}_\mathrm{Comb}=10.8$). \citet{Kass_Raftery_1995} argue that these large differences are a strong evidence against models with higher BIC-values. It is therefore concluded that the three scenarios with $e=0$ result in the best fits. Eventually, the combined data scenario with $e=0$ is selected which is consistent with G18 within the error. In section \ref{sec:228735255b} the results are compared in detail to those from G18.). 

\begin{figure}
\includegraphics[width=\columnwidth]{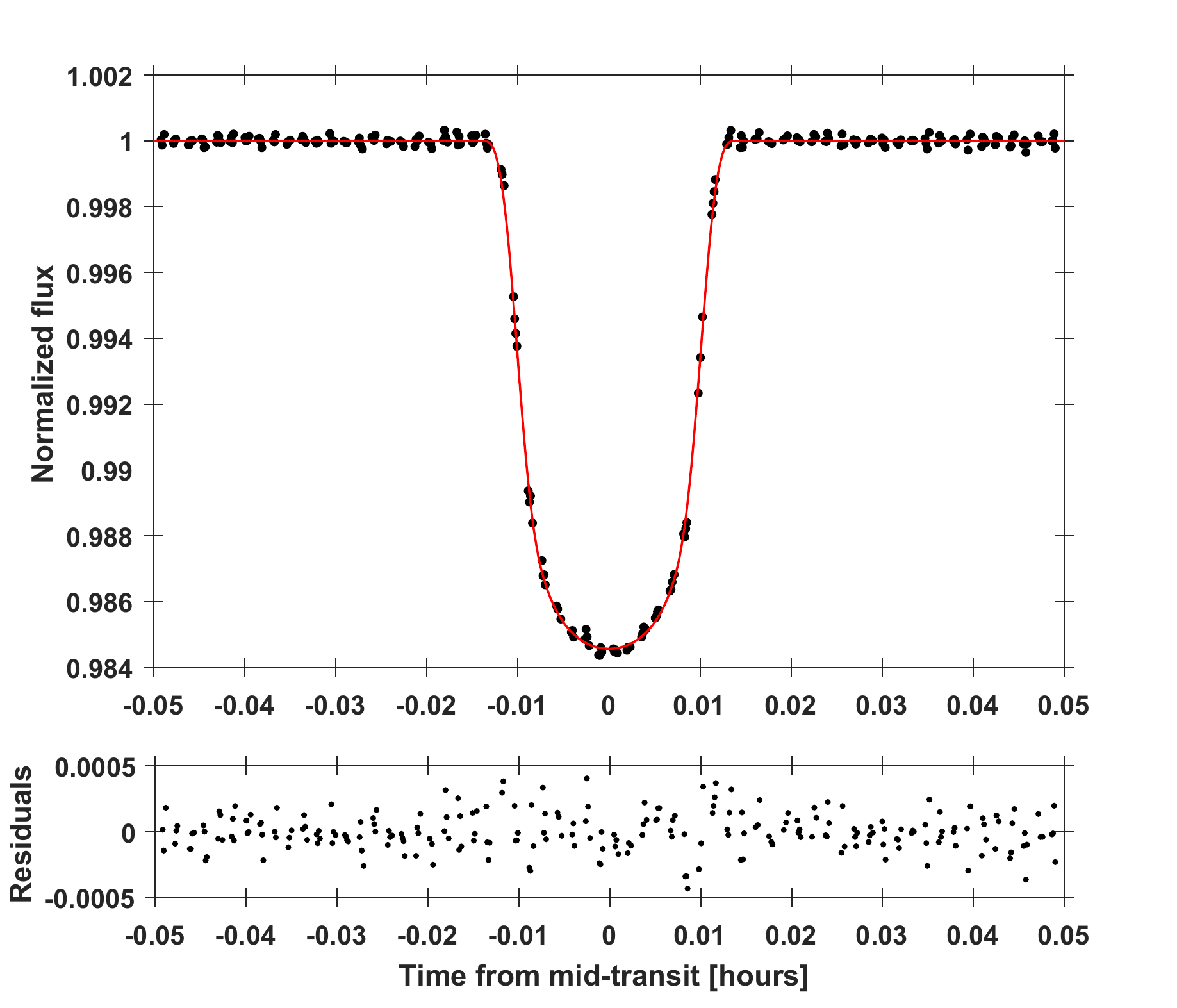}
\caption{\textit{Upper panel}: Best transit model for K2-140b (red solid line). Filled black circles are the folded light curve corrected for stellar variability. \textit{Lower panel}: Fit residuals.}
\label{fig:c10_lcfit}
\end{figure}

The $\chi^2$ of the best fit solution should be around 2 according to equation~\ref{eq:x2}. The final solution, however, has a $\chi^2=2.7$. The high $\chi^2$ of K2-140 means that a RV jitter has to be taken into account in the combined LC+RV fit. It turned out that the high $\chi^2$ is not caused by the RV but instead it is produced by in-transit variation due to spot-crossing events (see Fig. \ref{fig:c10_lcfit} bottom panel). This is also proved by a calculated RV jitter value of -1.3\,\ms\ for K2-140 which is too low to have an influence on the fit and was therefore neglected. This statement and the RV jitter value is also in good agreement with the value of -1\,\ms\ from G18. The standard deviation of the whole out-of-transit light curve is 140~ppm/30 min, and scaled to 6 hours, it is 37ppm/$\sqrt{6}$ which is comparable to the value found by \citet{vanderburg_johnson_2014} for this magnitude range.  

\begin{figure}
\includegraphics[width=\columnwidth]{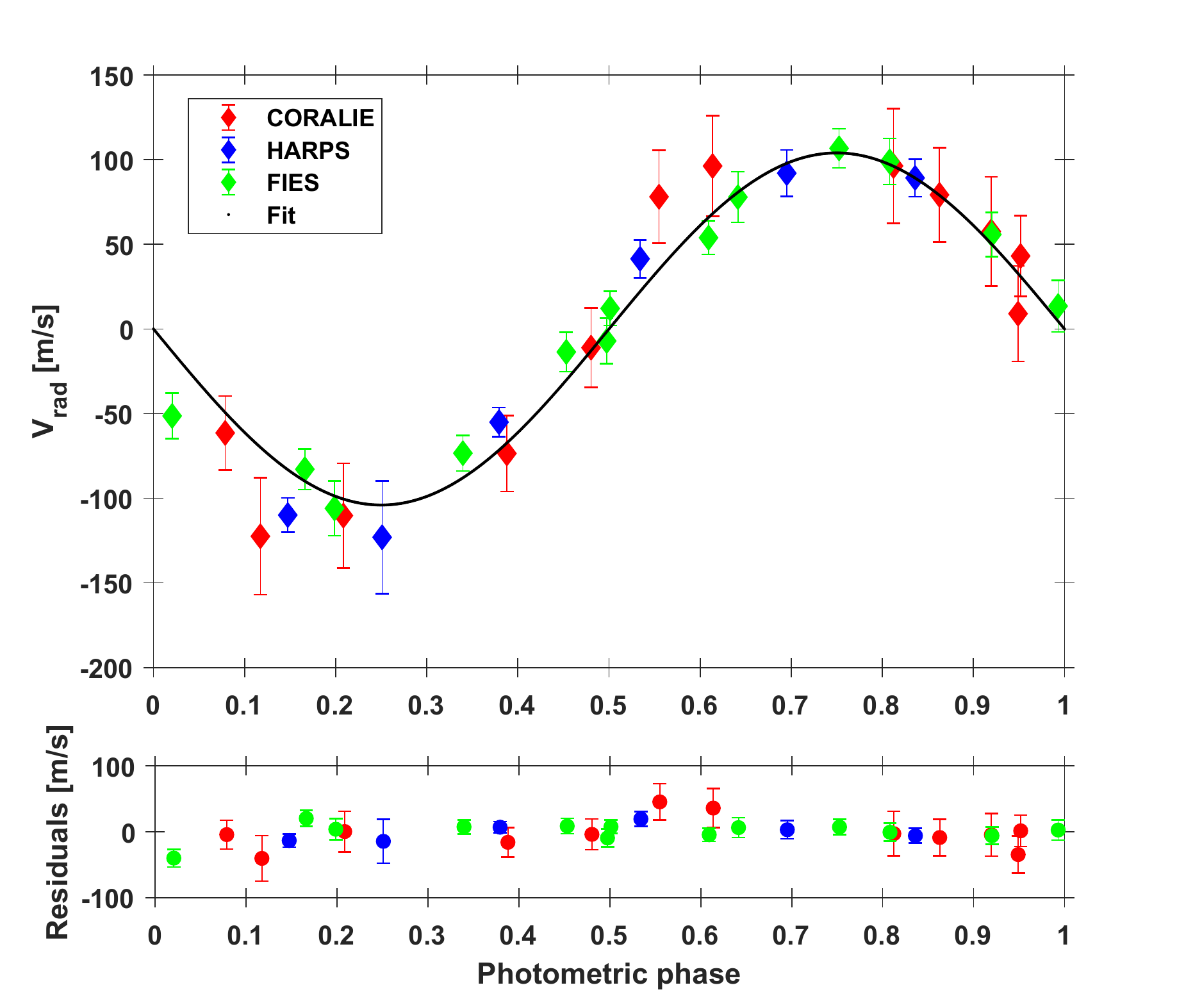}
\caption{\textit{Upper panel}: The best fit to the radial velocity data of K2-140 (black solid line). The CORALIE-, HARPS- and FIES-measurements are shown in red, blue and green, respectively. The CORALIE- and HARPS-data are from G18 while FIES data are from KESPRINT (this paper). \textit{Lower panel}: Fit residuals.}
\label{fig:c10_rvfit}
\end{figure}

The transit fit and RV fit are shown in Fig.~\ref{fig:c10_lcfit} and Fig.~\ref{fig:c10_rvfit}, respectively, for K2-140b. The derived parameters are listed in Table~\ref{tab:planparams_c5_9617}.

The orbital period of K2-140b and the exposure time applied by K2 (1726s) have a ratio of 1:328.1. This results in grouped data points in phase space (Fig.~\ref{fig:c10_lcfit}). Therefore, the first and the last contact (ingress and egress) are not well defined. As \citet{Smith_et_al_2018b} pointed out in that cases further constraints are needed because the transit duration (i.e. the scaled semi-major axis $a/R_\star$) is not well constrained, multiple degeneracies can be experienced. The degeneracies are between impact parameter and scaled semi-major axis, and between scaled semi-major axis and limb-darkening coefficients. \citet{Smith_et_al_2018b} pointed also out, that fixing, constraining or prioritizing the limb darkening coefficients might be misleading because there are significant differences between limb darkening coefficients calculated from 1D, 3D, plane-parallel or spherical symmetric stellar atmosphere models\footnote{Limb darkening coefficients can be over 1 or far from the \citet{Claret_Bloemen_cat_2011} values if one allows spherical symmetric stellar atmosphere models like \citet{Neilson_Lester_2013} did.}, and in addition, the observational checks are poor \citep[e.g.][]{csizmadia_et_al_2013,Neilson_Lester_2013}. Therefore it is a better strategy to adjust the limb darkening coefficients \citep{Morris_et_al_2018} and to prioritize the scaled semi-major axis calculated from the spectroscopically derived stellar density \citep{Smith_et_al_2018b}. The different treatment of the limb-darkening might explain the difference between the here presented values for $a/R_\star$ of 14.1 $\pm$ 0.2 calculated with TLCM and the values reported in G18 (12.7 $\pm$ 0.7), in \citet{Livingston_et_al_2018} ($15.1^{+0.1}_{-0.3}$), in \citet{mayo_et_al_2018} ($15.3^{+0.08}_{-1.9}$) and calculated with pyaneti ($15.2^{+0.1}_{-0.2}$). The difference between the data clumps is $\sim$ 0.001 in phase space (Fig.~\ref{fig:c10_lcfit}) which corresponds to $\pm$\,0.91 in $a/R_\star$. Taking this range into account all different $a/R_\star$ values agree within 1$\sigma$. Therefore, the different $a/R_\star$ values might arise by fact that the orbital period is close to a half-integer number of the exposure time causing that transit duration is less determinable for this case. 

\section{Discussion}
\subsection{K2-180b}

K2-180b was first reported as a planetary candidate by \citet{pope_et_al_2016} and recently validated as a planet by \citet{mayo_et_al_2018}. The planetary nature of the transiting signal is independently confirmed in this paper. The light curve analysis agrees well with both results. The combination of the \ktwo\ photometry with high-precision RV measurements yields a planetary mass of $M_\mathrm{p}=11.3\pm1.9$\,\Mearth\ and a radius of $R_\mathrm{p}=2.2\pm0.1$\,\Rearth, resulting in a bulk density of $\rho_\mathrm{p}=5.6\pm1.9$\,\gcm, suggesting that K2-180b is one of the densest mini-Neptune planet known so far. Particularly, K2-180b belongs to the so-called ``Hoptunes'' which are Neptunes (2\,\Rearth\,<\,$R_{\mathrm{p}}$\,<\,6\,\Rearth) with P\,<\,10\,days \citep{Dong_et_al_2018}. The different densities of mini-Neptune-size planets (radii between 2\,-\,4\,\Rearth) implies a wide range of possible compositions (Fig.~\ref{fig:zeng}), e.g. K2-110b has a density of $5.2\pm1.2$\,\gcm\,\citep{Osborn_et_al_2017}. K2-180b might have a ``rocky'' composition consisting mainly of magnesium silicate. However, a composition of a mixture up to 40\,\%\,H$_2$O and 60\,\%\,MgSiO$_3$ lies within the 1$\sigma$ uncertainty according to the theoretical models from \citet{Zeng_et_al_2016}. 

The region between 1\,-\,4\,\Rearth\ is of particular interest because of the so-called ``radius gap'' between 1.5 and 2 \Rearth\ for close-in planets \citep{fulton_et_al_2017,Van_Eylen_et_al_2018}. \citet{fulton_et_al_2017} found that only a few \kepler\ planets have radii between 1.5 and 2 \Rearth\ and orbital periods shorter than 100 days. Their conclusion was later confirmed using asteroseismic-derived stellar parameters by \citet{Van_Eylen_et_al_2018}. Such a radius gap has been predicted by models of photoevaporation \citep{Owen_wu_2013,lopez_fortney_2013}, and its observed slope \citep{Van_Eylen_et_al_2018} is consistent with such models \citep[e.g.][]{Owen_wu_2017,Jin_Mordasini_2018}. The main feature of these models is that a planet may lose its atmosphere due to stellar radiation. The radius gap marks the dividing line between super-Earths below the gap, which are stripped cores that have lost their entire atmosphere, and mini-Neptunes above the gap, which have held on to a gas envelope. Another possible mechanism has recently been suggested by \citet{Ginzburg_et_al_2018}, who proposed a core driven mechanism in which the luminosity of a cooling core may erode light envelopes and leave behind heavier envelopes. 

\cite{Van_Eylen_et_al_2018} measure the location of the gap as $\log R=m\log P+a$, and find $m=-0.09^{+0.02}_{-0.04}$ and $a=0.37^{+0.04}_{-0.02}$, for radii $R$ expressed in \Rearth\ and periods $P$ expressed in days. K2-180b has a radius of $2.2\pm0.1$\,\Rearth, and at an orbital period of $P=8.86$\,days, the gap is located at 1.9\,\Rearth. \citet{Fulton_petigura_2018} found a mass dependence of the radius gap. For low-mass stars the distribution is shifted to smaller sizes which is consistent with the fact that smaller stars produce smaller planetary cores \citep{Fulton_petigura_2018}. This makes K2-180b an interesting planet located just above the radius gap. Independent measurements of the radius by \citet{mayo_et_al_2018} and \citet{Petigura_et_al_2018}, which agree well within 1$\sigma$ and 2$\sigma$ with the planetary radius derived in this paper, find $R=2.41^{+0.21}_{-0.11}$\,\Rearth\ and $R_\mathrm{p}=4.4^{+5.9}_{-2.0}$\,\Rearth. According to the actual possible origins of the radius gap K2-180b is likely to have a gaseous envelope. The mass-radius diagram (Fig.~\ref{fig:zeng}) suggests also a relative massive core due to its density.

Intriguingly, given the low-metal content of its host star ([Fe/H]\,=\,$-0.65\pm0.10$), K2-180b is also one of the few mini-Neptune-size planets known to transit a metal-poor star. Known mini-Neptune-size planets orbiting metal-poor stars are e.g. HD\,175607 \citep{Mortier_et_al_2016,Faria_et_al_2016}. While a correlation between planetary mass and host star's metallicity is found for gas giants \citep[e.g.][]{Mortier_et_al_2012} the correlation for smaller planets is still investigated \citep{Wang_fischer_2015,Courcol_et_al_2016,Faria_et_al_2016,Mortier_et_al_2016,petigura_et_al_2018b}. \citet{Wang_fischer_2015} emphasized that the correlation between the occurrence and host star's metallicity seems to be weaker for terrestrial planets. Their statement was specified by \citet{Courcol_et_al_2016} who found this is true for Neptune-like planets (10\,\Mearth\,<\,$M_\mathrm{p}$\,<\,40\,\Mearth) but not for super-Earth planets ($M_\mathrm{p}$\,<\,10\,\Mearth).  With a mass of $M_\mathrm{p}=11.3\pm1.9$\,\Mearth\ K2-180b falls in the middle of both populations and is therefore of particular interest. It seems to exist also a desert of Neptune-like planets orbiting metal-poor stars ([Fe/H]\,<\,-0.5), also seen by \citet{petigura_et_al_2018b}. \citet{Dong_et_al_2018} found that Hoptunes are more common around metal-rich stars which highlights also K2-180b as a remarkable mini-Neptune-size planet. This facts puts K2-180b also in an outstanding position.  

These conclusions, however, depend on the photometric quality and on the accuracy of the stellar radius measurement. \citet{Petigura_et_al_2018} have done an independent check of the K2-180 light curve and stellar parameters. They obtained optical spectra using the High Resolution Echelle Spectrometer (HIRES) \citep{vogt_et_al_1994} on the Keck I 10 m telescope which are analyzed with the SpecMatch-Syn{\footnote{\url{https://github.com/petigura/specmatch-syn}} software code \citep{Petigura_2015}. \citet{mayo_et_al_2018} obtained also high-resolution spectra with the Tillinghast Reflector Echelle Spectrograph (TRES) at the Whipple Observatory to validate this system using the Validation of Exoplanet Signals using Probabilistic Algorithm (\texttt{VESPA}) \citep{Morton_2012,Morton_2015}. The stellar parameters from both studies agree within 1$\sigma$ with the stellar parameters derived in this paper (Table~\ref{tab:stellartab2}) except for \logg\ which agrees within 2$\sigma$. The derived stellar radii of $R_{\star}=0.65 \pm 0.02$\,\Rsun\ \citep{Petigura_et_al_2018}, $R_{\star}=0.68 \pm 0.02$\,\Rsun\ \citep{mayo_et_al_2018} and $R_{\star}=0.69 \pm 0.02$\,\Rsun\ from this study agree quite well (better than 7\,\%). Because of this good agreement, the differences in the derived planetary radii are rather caused by systematics resulting from the light curve detrending and/or noise modeling than by different stellar radius values. \citet{Petigura_et_al_2018} and \citet{mayo_et_al_2018} focused not on an individual system but on a list of candidates. Here, the red noise was included in the modeling procedure to obtain the planetary parameters at its best. The phase-folding can significantly affect accurate radius determination. Therefore, and due to the fact that the modeling includes RV measurements leads to better constrained parameters as in previously published papers. However, it would be worthwhile to observe the star again to obtain an additional light curve in order to refine the planetary radius (e.g. with CHEOPS\footnote{\url{http://cheops.unibe.ch}}).

\begin{figure}
\includegraphics[width=\columnwidth]{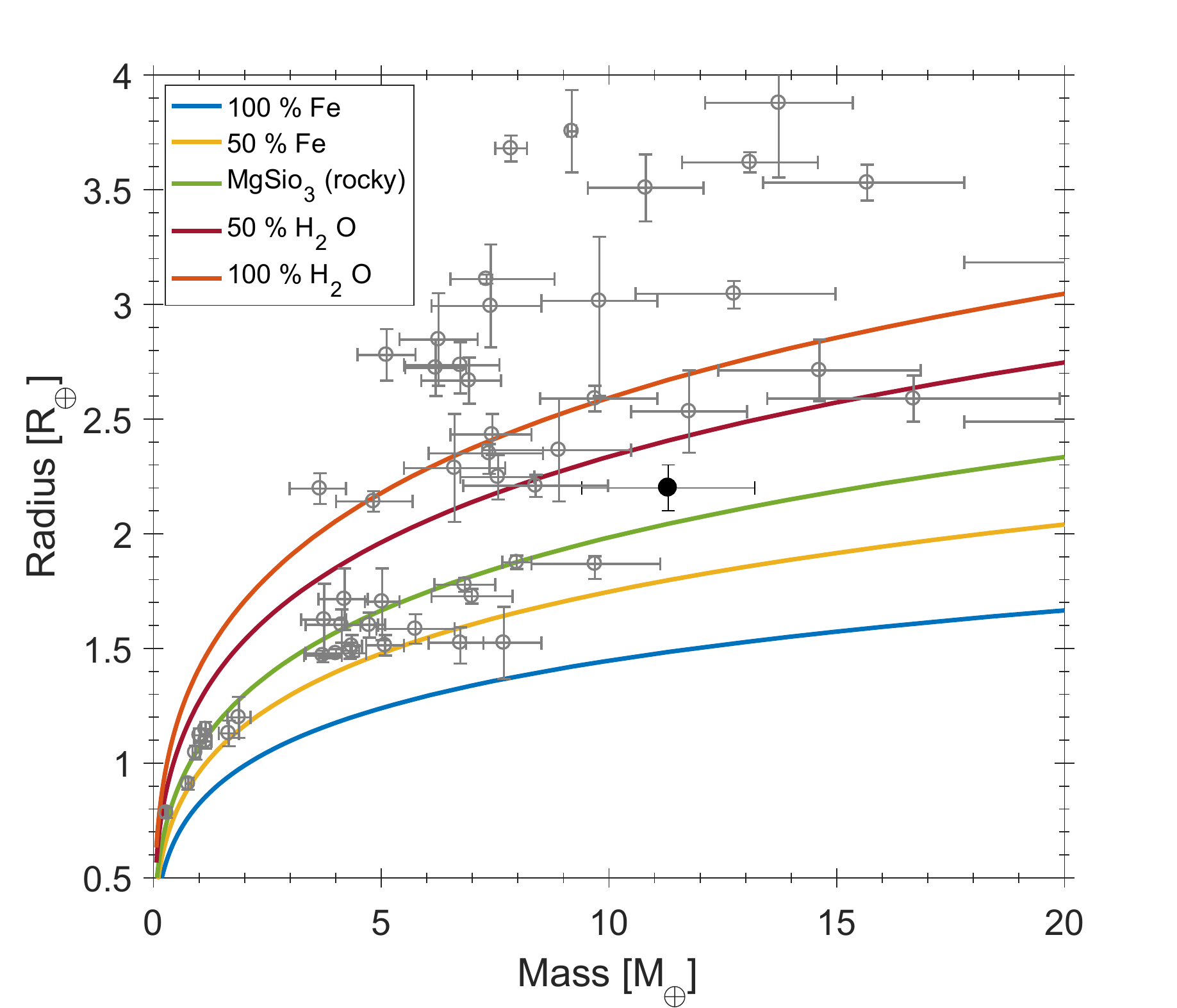}
\caption{Known planets with radii up to 4\,\Rearth (gray squares) with mass and radii accuracies better than 20\,\% (as of September 2018, TEPCat). K2-180b is marked with a black filled circle. The density models (solid lines) are curves of constant bulk density but varying composition \citep{Zeng_et_al_2016}.}
\label{fig:zeng}
\end{figure}

\subsection{K2-140b}
\label{sec:228735255b}

\begin{figure}
\includegraphics[width=\columnwidth]{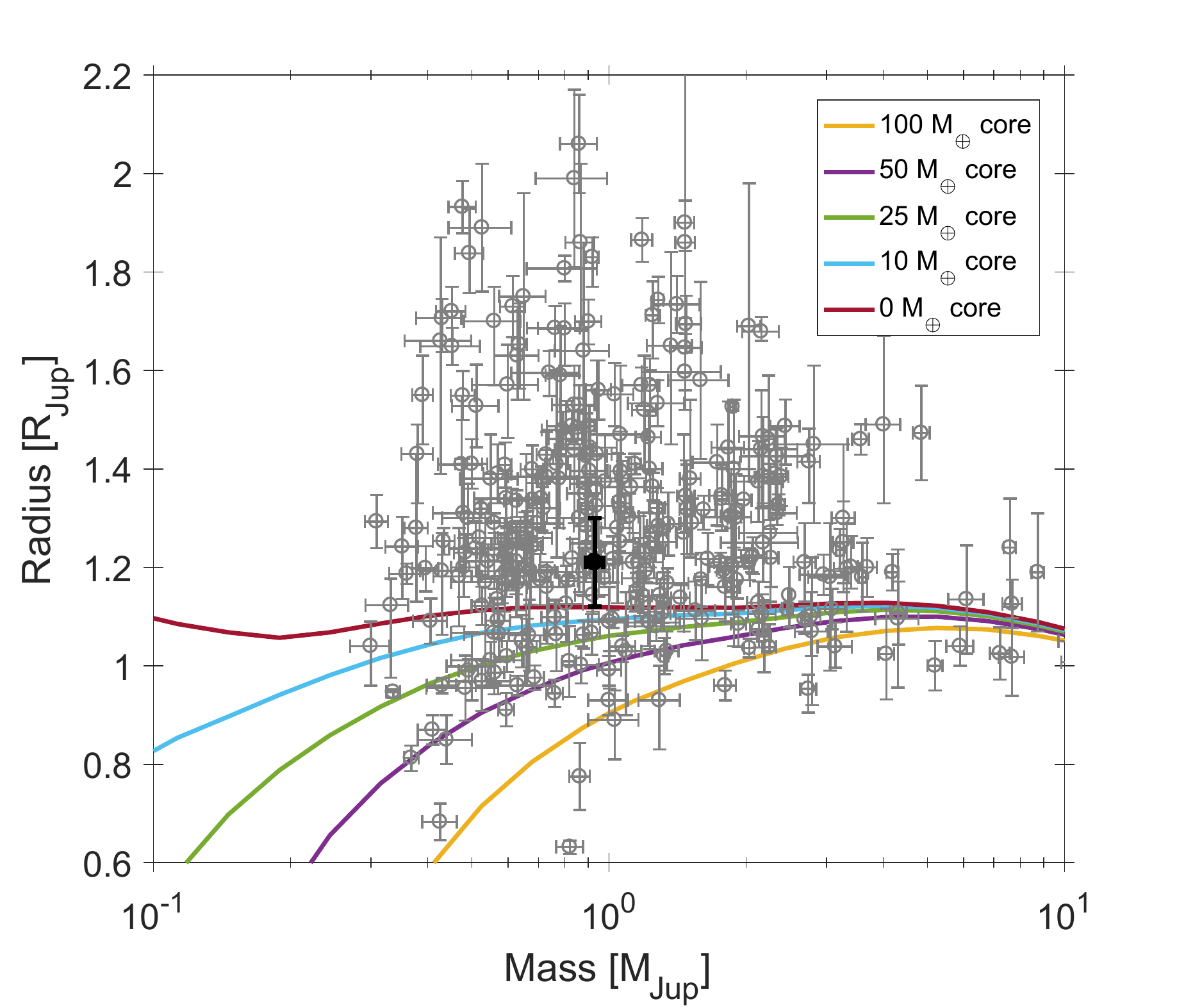}
\caption{Known hot Jupiters (gray squares) with mass and radius estimates better than 20\,\% (as of September 2018, TEPCat). K2-140b is marked with a black filled circle. The solid lines are the \citet{Fortney_et_al_2007} models for planet core masses of 0, 10, 25, 50 and 100 \Mearth.}
\label{fig:fortney}
\end{figure}

\citet{Livingston_et_al_2018} and \citet{mayo_et_al_2018} validated K2-140b with \texttt{VESPA} but came to different conclusions. While \citet{Livingston_et_al_2018} validated K2-140b as a planet, \citet{mayo_et_al_2018} reported it just as a candidate\footnote{They did not find a FFP value; see their notes in table 4.}. In contrast to this work, the mass estimate of K2-140b in \citet{Livingston_et_al_2018} was based on simple mass-radius-relation. It was not the intention to give accurate mass determinations, rather to select appropriate targets for follow-up observations. This highlights the importance of RV measurements to determine precise planetary masses. The planet mass derived by G18 using RV measurements, is in agreement with the presented results within 1$\sigma$. The main difference between the subsequent analysis and the parameters for K2-140b from G18 is the orbital eccentricity. An accurate estimate of this parameter is important since it affects the derived argument of pericenter, stellar mean density, semi-major axis and thus the derived equilibrium temperature. 

G18 found a non-zero eccentricity for K2-140b of $e=0.120^{+0.056}_{-0.046}$ using 12 CORALIE and 6 HARPS RV measurements. The analysis presented in this paper favors in contrast a circular orbit solution. In order to check whether this is a direct result from the additional 13 RV measurements the relation for an expected eccentricity error $\sigma(e)$ (Zakamska et al 2011) is used: 
\begin{equation} \label{eq:e}
\log \sigma(e) = 0.48 - 0.89 \log (K \cdot \sqrt{N} / \sigma_\mathrm{obs})
\end{equation}
where $K$ is the radial-velocity amplitude, $N$ is the number of RV observations and $\sigma_\mathrm{obs}$ is the average RV error of the observations. The G18 RV data with $K=111.2$\,\ms, $\sigma_\mathrm{obs}=23.6$\,\ms\ and $N=18$ do not constrain the orbit eccentricity better than $\sigma(e)=\pm0.21$. The KESPRINT RV data with $K=102$\,\ms, $\sigma_\mathrm{obs}=12.8$\,\ms\ and $N=13$ constrain the orbit eccentricity to $\sigma(e)= \pm 0.15$. Combining both data sets ($K=106$\,\ms, $\sigma_\mathrm{obs}=19.1$\,\ms\ and $N=31$) the orbit eccentricity is constrained to $\sigma(e)= \pm 0.14$. Thus it was concluded that the more accurate KESPRINT RV measurements improved the eccentricity detection. Moreover, a combined fit using the light curve and the RV data achieves better accuracy in the eccentricity. The eccentricity found by G18 might be driven by the uncertainties of the individual RV measurements and their distribution along the RV curve. 

The significance of their result was here tested statistically by Monte Carlo simulations. One hundred thousand simulated RV datasets were created by sampling a best fitting circular orbit solution at the timestamps of the CORALIE and HARPS observations. Gaussian noise at the same level of the G18 measurements was added. A least-squares fit to these simulated data allowed also an eccentric solution. There is a probability of $\sim$12\,\% that an eccentric solution with $e\ge0.12$ may be found if the true orbit is actually circular. The same analysis was repeated by combining the FIES RV data with those from G18. Once again, simulated data created from the best fitting circular orbit model can mimic an eccentric orbit with a probability of 6\,\%. The new analysis shows that the previously claimed eccentric orbit is not statistically significant. In contrast, the BIC which distinguishes between an eccentric or circular solution prefers the circular orbit. The results highlight the difficulty to distinguish between an elliptical and circular orbit if the eccentricity is very small. In general, it is difficult to distinguish between a slightly eccentric orbit and a circular orbit in a light curve or RV data. 

The mass-radius diagram for hot Jupiters with masses and radii known to a precision better than 20\,\% is shown in Fig.~\ref{fig:fortney} ($P_\mathrm{orb}$\,<\,10\,days and $M_\mathrm{p}>0.3$\,\Mjup). K2-140b joins the group of well characterized hot Jupiters with a mean density of $\rho_\mathrm{p}=0.66\pm0.18$\,\gcm. These value agree with the mass-density-relationship from \citet{Hatzes_Rauer_2015} for giant planets, with K2-140b belonging to the sub-group of low-mass giant planets. K2-140b could have a core mass up to $\sim$8\,\Mearth\ after \citet{Fortney_et_al_2007}. This composition depends, however, again on the accuracy of the stellar parameters. The errors in stellar parameters reported in Table~\ref{tab:stellartab2} are the actual measurement errors. Stellar masses and radii in Table~\ref{tab:stellartab2} are based on the use of a particular model and their errors do not reflect the uncertainty of stellar models. It is only with the systematic observations of asteroseismic signals that more precise results are to be expected in the near future \citep{Rauer_et_al_2014}. 

The brightness of the host star and the transit depth makes K2-140 appropriate target for transit spectroscopy. Most transit spectroscopy to date has been performed on short period planets. Planet K2-140b could be an interesting target when investigating the change of atmospheric properties from short period hot Jupiters to longer period hot Jupiters.

\subsubsection{Tidal dynamics}

\begin{table*}
\centering
\caption{Tidal interaction parameters for K2-180, K2-140 and the CoRoT-21 system.}
\label{tab:tidal}
\begin{tabular}{lrrr}
 \hline
 \hline
 \noalign{\smallskip}
  Parameter & K2-180 & K2-140 & CoRoT-21 \\
  \noalign{\smallskip}
  \hline
  \noalign{\smallskip}
  $a$ [AU]  & 0.075 $\pm$ 0.001 & 0.068 $\pm$ 0.001 & 0.0417 $\pm$ 0.0011\\
  $a_\mathrm{sync}$ [AU] & 0.11 $\pm$ 0.03 & 0.11 $\pm$ 0.02 & 0.092 $\pm$ 0.019\\
  $D_\mathrm{p}$ [m$^{2}$\,s$^{-2}$] & 577 $\pm$ 104 & 47215 $\pm$ 6881 & (1.7 $\pm$ 0.5) $\cdot10^{6}$ \\
  $D_\star$ [m$^{2}$\,s$^{-2}$] & 10414 $\pm$ 1296 & 691500 $\pm$ 115673 & 4.6 $\pm$ 0.6) $\cdot10^{6}$\\
  $D_\star/D_p$ & 18.0 $\pm$ 4.0 & 14.7 $\pm$ 3.2 & 2.76 $\pm$ 0.82\\
  $F$ [$10^{56}$\,kg$^{0.5}$\,m$^5$] & 0.015 $\pm$  0.003 & 2.8 $\pm$ 0.9 & 124 $\pm$ 69\\
  $T$ [Gyr] & 26.1 $\pm$ 3.1 & 11.2 $\pm$ 2.0 & 4.9\\
  $\tau$ [Gyr] & 16.6 $\pm$ 6.4 & 1.4 $\pm$ 3.9 & 0.8 $\pm$ 0.5\\
  $a_\mathrm{roche}$ [AU] & 0.0063 $\pm$ 0.0005 & 0.014 $\pm$ 0.001 & 0.0127 $\pm$ 0.0008\\
  $a_\mathrm{crit}$ [$Q_\star/k_2=10^6$] & 0.024 $\pm$ 0.001 & 0.04 $\pm$ 0.02 & 0.060 $\pm$ 0.006 \\
  $a_\mathrm{crit}$ [$Q_\star/k_2=10^7$] & 0.017 $\pm$ 0.001 & 0.03 $\pm$ 0.01 & 0.042 $\pm$ 0.004\\
  $a_\mathrm{crit}$ [$Q_\star/k_2=10^8$] & 0.0118 $\pm$ 0.0007 & 0.019 $\pm$ 0.008 & 0.030 $\pm$ 0.002\\
  $a_\mathrm{crit}$ [$Q_\star/k_2=10^9$] & 0.0085 $\pm$ 0.0004 & 0.015 $\pm$ 0.003 & 0.021 $\pm$ 0.002\\
 \noalign{\smallskip}
  \hline
  \hline
 \end{tabular}
\end{table*}

If K2-140b experiences significant tidal interaction with its host star was also investigated by computing four principle parameters:
\begin{enumerate}
\item The synchronous orbit: a planet in an orbit about a star may show orbital decay if the planetary orbit is within the synchronous orbital radius $a_\mathrm{sync}$ defined by the stellar rotation rate $\Omega_\star$:
\begin{equation}
a_\mathrm{sync}= \left( \frac{G\cdot\left(M_\star+M_\mathrm{p}\right)}{\Omega_\star^{2}}\right)^{\frac{1}{3}},
\end{equation}
with $G$ as the gravitational constant. K2-140b is within the synchronous orbit about its host star. 
\item The planetary Doodson constant $D_\mathrm{p}$ which describes the magnitude of tidal forces acting from the planet on the star which may likely result in orbital decay and stellar rotation spin-up \citep{paetzold_et_al_2004}:
\begin{equation}
D_\mathrm{p}=\frac{3GM_\mathrm{p}R_\star^{2}}{4a^{3}}.
\end{equation}
\item The stellar property factor $F$ which describes the magnitude of orbital decay by stellar and planetary parameters 
\citep{paetzold_rauer_2002}: 
\begin{equation}
F=\frac{M_\mathrm{p}R_\star^{5}}{\sqrt{M_\star}}.
\end{equation}
\item The critical radius $a_\mathrm{crit}$: The planetary orbit decays fully within the remaining life time of the star with respect to a certain dissipation constant $\frac{Q_\star}{k_\mathrm{2}}$: 
\begin{equation}
a_\mathrm{crit}\geq\left(\frac{13}{2}\tau A\left(\frac{Q_\star}{k_2}\right)^{-1}+a_\mathrm{roche}^{\frac{13}
{2}}\right)^{\frac{2}{13}},
\end{equation}
where $\tau=T-T_\mathrm{age}$ is the remaining lifetime of the star, $T_\mathrm{age}$ is the age of the star and $T$ is the life time of the star estimated from \citet{Prialnik_2000}:
\begin{equation}
T=10^{10}\left(\frac{M_\star}{M_\odot}\right)^{-2.8} \mathrm{years}. 
\end{equation}
$A$ is a constant of stellar and planetary parameters \citep{carone_2012}:
\begin{equation}
A=3\frac{M_\mathrm{p}}{M_\star}R_\star^{5} \sqrt{G\cdot\left(M_\star+ M_\mathrm{p}\right)}.
\end{equation}
If the orbit decays, the planet is considered to be destroyed if it enters the stellar Roche zone $a_\mathrm{roche}$:
\begin{equation}
a_\mathrm{roche}=BR_\mathrm{p}\left(\frac{M_\star}{M_\mathrm{p}}\right)^{\frac{1}{3}},
\end{equation}
with $B = 2.44$ and $B = 1.44$ for gas giants and rocky planets, respectively \citep{Sharma_2009}.
\end{enumerate}
These four parameters are listed in Table~\ref{tab:tidal} for K2-140 compared to CoRoT-21 \citep{paetzold_et_al_2012}, a system with strong tidal interaction. Although K2-140b is within the synchronous orbit of its host star, it is well outside of critical radius for any reasonable stellar dissipation constant. The stellar Roche zone will not be entered within the remaining life time of the host star. This is also reflected in the small Doodson constant $D_\mathrm{p}$ and property factor $F$, orders of magnitude smaller than those of the CoRoT-21 system. The Neptune-sized planet, K2-180b, was also investigated and shows also no tidal interaction (Table~\ref{tab:tidal})

\section{Conclusions}
The KESPRINT consortium confirms, using high-resolution imaging and RV measurements, K2-180b, a mini-Neptune-size planet in a 8.87-day orbit that was reported as a planetary candidate by \citet{pope_et_al_2016} and recently validated by \citet{mayo_et_al_2018}. K2-180b has a mass of $M_\mathrm{p}=11.3\pm1.9$\,\Mearth\ and a radius of $R_\mathrm{p}=2.2\pm0.1$\,\Rearth, yielding a mean density of $\rho_\mathrm{p}=5.6\pm1.9$\,\gcm. With a radius of 2.2\,\Rearth, K2-180b lies slightly above radius valley, i.e., the bimodal distribution of planetary radii, with super-Earth and mini-Neptunes separated at $\sim$1.9\,\Rearth\ \citep{fulton_et_al_2017,Van_Eylen_et_al_2018} suggesting that K2-180b has a gaseous envelope. According to the theoretical models from \citet{Zeng_et_al_2016} K2-180b might have a ``rocky'' composition with a relative massive core for its size. The detection of the radius gap base on statistical analyses of \kepler\ planets for which precise mass measurements are not always available. This highlights the importance for accurate mass measurements for planets in the range from $1-4$\,\Rearth\ to further investigate the origin of the radius gap. K2-180 is also relatively unique in its low metallicity ([Fe/H]\,=\,$-0.65\pm0.10$) among transiting Neptune-size planets \citep{Wang_fischer_2015,Courcol_et_al_2016}. Both facts, a dense mini-Neptune-size planet which lies just above the radius gap transiting a metal-poor star, make it an ideal target for the upcoming CHEOPS and ARIEL missions. Separately, it is interesting that K2-180b is orbiting a metal-poor star, since \citet{Dong_et_al_2018} found such planets to be most common around metal-rich stars. 

K2-140b was previously confirmed by G18 who used RV measurements to determine the mass and the orbital eccentricity. In the present paper, the re-determination of the K2-140 system's parameters was performed including 13 additional FIES RV measurements, enabling a more precise derivation of the properties of the system. The new data constrain, in particular, the eccentricity better. The new analysis shows that the previously claimed eccentric orbit is not statistically significant. Given the current data-set, there is a 6\% probability that the non-zero eccentricity might arise if the underlying orbit is actually circular. The results highlight the difficulty to distinguish between an elliptical and circular orbit if the eccentricity is very small. Knowing the eccentricity is important e.g. for understanding the formation and evolution of a planetary system.  

The orbital period of K2-140b is a multiple of the exposure time applied by K2 resulting in grouped data points in phase space. One consequence is that the first and the last contact are not well defined and therefore the transit duration is not well constrained, yielding to degeneracies and to a larger than usual  uncertainty in the scaled semi-major axis $a/R_\star$. Therefore it is a better strategy to adjust the limb darkening coefficients \citep{Morris_et_al_2018} and to prioritize the scaled semi-major axis calculated from the stellar density based on spectroscopic \teff, \logg, and [Fe/H] \citep{Smith_et_al_2018b}.

With a mass of $M_\mathrm{p}=0.93\pm0.04$\,\Mjup\ and a radius of $R_\mathrm{p}=1.21\pm0.09$\,\Rjup, K2-140b has a mean density of $\rho_\mathrm{p}=0.66\pm0.13$\,\gcm, which follows the mass-density-relationship described by \citet{Hatzes_Rauer_2015}. According to the evolutionary models from \citet{Fortney_et_al_2007}, K2-140b might have a core's mass of up to $\sim$8\,\Mearth. Although K2-140b is within the synchronous orbit of its host star, it is well outside of the critical orbital radius for any reasonable stellar dissipation constant. The stellar Roche zone will also not be entered within the remaining lifetime of the host star. Therefore, K2-140b does not experience significant tidal interaction. The brightness of the host star and the transit depth make K2-140b a good target for transmission spectroscopy. A detailed calculation of the atmospheric characteristics was recently performed in \citet{Livingston_et_al_2018}. Transit spectroscopy has been so far conducted on short period planets. It would be therefore interesting to observe K2-140b to study how the atmospheric properties change from short period hot Jupiters ($P_\mathrm{orb}<3$\,days) to hot Jupiters with longer periods.    

Although different, both planets are valuable members of the sample of well-characterized systems needed to understand exoplanet diversity. While K2-140b is a "typical" hot Jupiter, K2-180b is a relative unique Hoptune based on its high density in the radius-mass regime of the detected sample of the Neptune-size population.

\section*{Acknowledgements}

We are very grateful to the NOT and TNG staff members for their unique and superb support during the observations. The research leading to these results has received funding from the European Union Seventh Framework Programme (FP7/2013-2016) under grant agreement No. 312430 (OPTICON). Based on observations obtained with the Nordic Optical Telescope (NOT), operated on the island of La Palma jointly by Denmark, Finland, Iceland, Norway, and Sweden, in the Spanish Observatorio del Roque de los Muchachos (ORM) of the Instituto de Astrof\'isica de Canarias (IAC). The WIYN/NESSI observations were conducted as part of NOAO observing program ID 2017A-0377 (P.I. Livingston). Data presented herein were obtained at the WIYN Observatory from telescope time allocated to NN-EXPLORE through the scientific partnership of the National Aeronautics and Space Administration, the National Science Foundation, and the National Optical Astronomy Observatory. This work was supported by a NASA WIYN PI Data Award, administered by the NASA Exoplanet Science Institute. NESSI was funded by the NASA Exoplanet Exploration Program and the NASA Ames Research Center. NESSI was built at the Ames Research Center by Steve B. Howell, Nic Scott, Elliott P. Horch, and Emmett Quigley. This work has made use of data from the European Space Agency (ESA) mission \gaia\ (\url{https://www.cosmos.esa.int/gaia}), processed by the \gaia\ Data Processing and Analysis Consortium (DPAC, \url{https://www.cosmos.esa.int/web/gaia/dpac/consortium}). Funding for the DPAC has been provided by national institutions, in particular the institutions participating in the \gaia\ Multilateral Agreement. This publication makes use of data products from the Two Micron All Sky Survey (2Mass), which is a joint project of the University of Massachusetts and the Infrared Processing and Analysis Center/California Institute of Technology, funded by the National Aeronautics and Space Administration and the National Science Foundation. Funding for the Stellar Astrophysics Centre is provided by The Danish National Research Foundation (Grant agreement no.: DNRF106).  J.~Korth, S.~Grziwa, M.~P\"atzold, Sz.~Csizmadia, A.~P.~Hatzes, and H.~Rauer acknowledge support by DFG grants PA525/18-1, PA525/19-1, PA525/20-1, HA 3279/12-1 and RA 714/14-1 within the DFG Schwerpunkt SPP 1992, Exploring the Diversity of Extrasolar Planets. Sz.~Csizmadia acknowledges the Hungarian OTKA Grant K113117. D.~Gandolfi acknowledges the financial support of the \emph{Programma Giovani Ricercatori -- Rita Levi Montalcini -- Rientro dei Cervelli (2012)} awarded by the Italian Ministry of Education, Universities and Research (MIUR). M.~Fridlund and C.~M.~Persson gratefully acknowledge the support of the Swedish National Space Board. C.~E.~Petrillo and C.~Tortora are supported through an NWO-VICI grant (project number 639.043.308). This work is partly supported by JSPS KAKENHI Grant Number JP18H01265.




\bibliographystyle{mnras}
\bibliography{references3} 



\bsp	
\label{lastpage}
\end{document}